\documentclass[namedreferences]{solarphysics}
\usepackage[optionalrh,solaromanenum]{spr-sola-addons} 
\usepackage{graphicx}                    
\usepackage{xcolor}                       
\usepackage{textgreek}                       

\usepackage{float}
\usepackage{soul}                      

\newcommand{\arcsec}{\ensuremath{^{\prime\prime}}}
\newcommand{\arcmin}{\ensuremath{^{\prime}}}

\begin{document}

\begin{article}

\begin{opening}

\title{The High-Resolution Coronal Imager, Flight 2.1}

%
\author[addressref={aff1},corref,email={laurel.rachmeler@nasa.gov}]{\inits{L.A.}\fnm{Laurel~A.}~\lnm{Rachmeler}}
\author[addressref={aff1},corref,email={amy.r.winebarger@nasa.gov}]{\inits{A.R.}\fnm{Amy~R.}~\lnm{Winebarger}}
\author[addressref={aff1},corref]{\inits{S.L.}\fnm{Sabrina~L.}~\lnm{Savage}}
\author[addressref={aff2},corref]{\inits{L.}\fnm{Leon}~\lnm{Golub}\orcid{0000-0001-9638-3082}}
\author[addressref={aff1},corref]{\inits{K.}\fnm{Ken}~\lnm{Kobayashi}}
\author[addressref={aff12},corref]{\inits{G.D.}\fnm{Genevieve~D.}~\lnm{Vigil}}
\author[addressref={aff3},corref]{\inits{D.H.}\fnm{David~H.}~\lnm{Brooks}}
\author[addressref={aff4},corref]{\inits{J.W.}\fnm{Jonathan~W.}~\lnm{Cirtain}}
\author[addressref={aff5,aff16,aff17},corref]{\inits{B.}\fnm{Bart}~\lnm{De~Pontieu}\orcid{0000-0002-8370-952X}}
\author[addressref={aff1},corref]{\inits{D.E.}\fnm{David~E.}~\lnm{McKenzie}}
\author[addressref={aff7},corref]{\inits{R.J.}\fnm{Richard~J.}~\lnm{Morton}\orcid{0000-0001-5678-9002}}
\author[addressref={aff8},corref]{\inits{H.}\fnm{Hardi}~\lnm{Peter}\orcid{0000-0001-9921-0937}}
\author[addressref={aff2},corref]{\inits{P.}\fnm{Paola}~\lnm{Testa}}
\author[addressref={aff5,aff9},corref]{\inits{S.K.}\fnm{Sanjiv~K.}~\lnm{Tiwari}\orcid{0000-0001-7817-2978}}
\author[addressref={aff10},corref]{\inits{R.W.}\fnm{Robert~W.}~\lnm{Walsh}}
\author[addressref={aff11},corref]{\inits{H.P.}\fnm{Harry~P.}~\lnm{Warren}\orcid{0000-0001-6102-6851}}
\author[addressref={aff12},corref]{\inits{C.}\fnm{Caroline}~\lnm{Alexander}}
\author[addressref={aff10},corref]{\inits{D.}\fnm{Darren}~\lnm{Ansell}}
\author[addressref={aff1},corref]{\inits{B.L.}\fnm{Brent~L.}~\lnm{Beabout}}
\author[addressref={aff1},corref]{\inits{D.L.}\fnm{Dyana~L.}~\lnm{Beabout}}
\author[addressref={aff12},corref]{\inits{C.W.}\fnm{Christian~W.}~\lnm{Bethge}}
\author[addressref={aff1},corref]{\inits{P.R.}\fnm{Patrick~R.}~\lnm{Champey}}
\author[addressref={aff2},corref]{\inits{P.N.}\fnm{Peter~N.}~\lnm{Cheimets}}
\author[addressref={aff1},corref]{\inits{M.A.}\fnm{Mark~A.}~\lnm{Cooper}}
\author[addressref={aff1},corref]{\inits{H.K.}\fnm{Helen~K.}~\lnm{Creel}}
\author[addressref={aff2},corref]{\inits{R.}\fnm{Richard}~\lnm{Gates}}
\author[addressref={aff1},corref]{\inits{C.F.}\fnm{Carlos}~\lnm{Gomez}}
\author[addressref={aff1},corref]{\inits{R.}\fnm{Anthony}~\lnm{Guillory}}
\author[addressref={aff1},corref]{\inits{H.}\fnm{Harlan}~\lnm{Haight}}
\author[addressref={aff1},corref]{\inits{W.D.}\fnm{William~D.}~\lnm{Hogue}}
\author[addressref={aff1},corref]{\inits{T.}\fnm{Todd}~\lnm{Holloway}}
\author[addressref={aff1},corref]{\inits{D.W.}\fnm{David~W.}~\lnm{Hyde}}
\author[addressref={aff10},corref]{\inits{R.}\fnm{Richard}~\lnm{Kenyon}}
\author[addressref={aff1},corref]{\inits{J.N.}\fnm{Joseph~N.}~\lnm{Marshall}}
\author[addressref={aff1},corref]{\inits{J.E.}\fnm{Jeff~E.}~\lnm{McCracken}}
\author[addressref={aff2},corref]{\inits{K.}\fnm{Kenneth}~\lnm{McCracken}}
\author[addressref={aff1},corref]{\inits{K.O.}\fnm{Karen~O.}~\lnm{Mitchell}}
\author[addressref={aff2},corref]{\inits{M.}\fnm{Mark}~\lnm{Ordway}}
\author[addressref={aff1},corref]{\inits{T.}\fnm{Tim}~\lnm{Owen}}
\author[addressref={aff12},corref]{\inits{J.}\fnm{Jagan}~\lnm{Ranganathan}}
\author[addressref={aff1},corref]{\inits{B.A.}\fnm{Bryan~A.}~\lnm{Robertson}}
\author[addressref={aff1},corref]{\inits{M.J.}\fnm{M.~Janie}~\lnm{Payne}}
\author[addressref={aff2},corref]{\inits{W.}\fnm{William}~\lnm{Podgorski}}
\author[addressref={aff1},corref]{\inits{J.}\fnm{Jonathan}~\lnm{Pryor}}
\author[addressref={aff2},corref]{\inits{J.}\fnm{Jenna}~\lnm{Samra}}
\author[addressref={aff15},corref]{\inits{M.D.}\fnm{Mark~D.}~\lnm{Sloan}}
\author[addressref={aff1},corref]{\inits{H.A.}\fnm{Howard~A.}~\lnm{Soohoo}}
\author[addressref={aff1},corref]{\inits{D.B.}\fnm{D.~Brandon}~\lnm{Steele}}
\author[addressref={aff1},corref]{\inits{F.V.}\fnm{Furman~V.}~\lnm{Thompson}}
\author[addressref={aff1},corref]{\inits{G.S.}\fnm{Gary~S.}~\lnm{Thornton}}
\author[addressref={aff10},corref]{\inits{B.}\fnm{Benjamin}~\lnm{Watkinson}}
\author[addressref={aff13},corref]{\inits{D.}\fnm{David}~\lnm{Windt}}

\runningauthor{Rachmeler et al.}
\runningtitle{Hi-C\,2.1}


\address[id=aff1]{NASA Marshall Space Flight Center, 320 Sparkman Dr, Huntsville, AL 35805, USA}
\address[id=aff2]{Harvard-Smithsonian Center for Astrophysics, 60 Garden Street, Cambridge, MA 02138, USA}
\address[id=aff12]{Universities Space Research Association, 320 Sparkman Dr., Huntsville, AL  35805, USA}
\address[id=aff3]{College of Science, George Mason University, 4400 University Drive, Fairfax, VA 22030, USA}
\address[id=aff4]{BWX Technologies, Inc., 800 Main St. Suite 400, Lynchburg, VA 24504, USA}
\address[id=aff5]{Lockheed Martin Solar and Astrophysics Laboratory, 3251 Hanover st., Org. A021S, Bldg.252, Palo Alto, CA, 94304, USA}
\address[id=aff16]{Rosseland Centre for Solar Physics, University of Oslo, P.O. Box 1029 Blindern, NO-0315, Oslo, Norway}
\address[id=aff17]{Institute of Theoretical Astrophysics, University of Oslo, P.O. Box 1029 Blindern, NO-0315, Oslo, Norway}
\address[id=aff7]{Department of Mathematics, Physics and Electrical Engineering, Northumbria University, Newcastle Upon Tyne, NE1 8ST, UK}
\address[id=aff8]{Max Planck Institute for Solar System Research, 37077 G\"{o}ttingen, Germany}
\address[id=aff9]{Bay Area Environmental Research Institute, NASA Research Park, Moffett
Field, CA 94035, USA}
\address[id=aff10]{University of Central Lancashire, Preston, PR1 2HE, UK}
\address[id=aff11]{Space Science Division, Naval Research Laboratory, Washington, DC 20375 USA}
\address[id=aff15]{Jacobs, 620 Discovery Dr., Suite 140, Huntsville, AL 35806}
\address[id=aff13]{Reflective X-ray Optics LLC, 425 Riverside Dr., \#16G, New York, NY 10025}

\begin{abstract}

  The third flight of the High-Resolution Coronal Imager
    (Hi-C\,2.1) occurred on May 29, 2018, the Sounding Rocket was launched from White Sands Missile Range
    in New Mexico. The instrument has been modified from its original
    configuration (Hi-C\,1) to observe the solar corona in a passband that peaks
    near 172\,\AA, and uses a new, custom-built low-noise
    camera. The instrument targeted Active Region 12712, and captured
    78 images at a cadence of 4.4\,sec (18:56:22 - 19:01:57~UT; 5\,min and 35\,sec observing time). The image spatial resolution varies due to quasi-periodic motion blur from the rocket; sharp images contain resolved features of at least 0.47\,arcsec. There are coordinated observations from multiple ground- and space-based telescopes providing an unprecedented opportunity to observe the mass and energy coupling between the chromosphere and the corona. Details of the instrument and the data set are presented in this paper. 

\end{abstract}

%

\end{opening}

%

\section{Introduction} \label{sec:intro}

The High-resolution Coronal Imager (Hi-C) has been launched three times from White Sands Missile Range (WSMR). The first launch, Hi-C\,1, occurred on July 11, 2012 \citep{kobayashi2014}. During the 345 seconds of data acquisition, Hi-C\,1 obtained the highest spatial resolution and highest cadence images of the extreme ultraviolet (EUV) solar corona of an active region (AR) ever achieved. Those few minutes of $193$\,\AA\ data have thus far generated over 25
refereed publications, including the first ever observation of coronal braiding and associated energy release \citep{2013Natur.493..501C}. Other results include evidence of nanoflaring in AR coronal (braided) magnetic loops \citep{2013ApJ...771...21W,2014ApJ...780..102T,2014ApJ...795L..24T,2017ApJ...837..108P} and in surrounding moss \citep{2013ApJ...770L...1T}, sub-structure in coronal loops \citep{2013A&A...556A.104P,2013ApJ...772L..19B}, small-scale ``sparkling" dynamic bright dots in the AR moss \citep{2014ApJ...784..134R}, counter-streaming flows in filaments \citep{2013ApJ...775L..32A}, and moving bright dots and jets in the sunspot penumbra \citep{2016ApJ...822...35A,2016ApJ...816...92T}.

Following the success of the first flight, Hi-C was modified to observe in a different wavelength to study the mass and energy coupling between the chromosphere and the corona (Hi-C\,2). For this objective, the wavelength of the passband was changed to $172$\,\AA\ (Fe \textsc{ix/x}) and a new custom-built, low-noise camera was installed. This passband samples plasma at a relatively cool coronal or transition region temperature of $\sim1$\,MK. Hi-C\,2 was designed to study two scientific questions: are there coronal counterparts to type II spicules; and what is the relationship between chromospheric and coronal heating in active region cores? The cooler temperatures and high temporal and spatial resolution allow the data to be combined with co-observations made by the Interface Region Imaging Spectrograph \citep[IRIS;][]{2014SoPh..289.2733D} to study this connection between the chromosphere and the corona. 

The type-II spicules inject chromospheric plasma upwards at velocities of order $50-100$\,km/s and are often seen to fade as they are heated out of the chromospheric passbands \citep{2007PASJ...59S.655D}. While they may play an important role in mass and energy injected into the corona, their coronal counterparts are difficult to study because they are faint, small (diameters $<0.5$\,arcsec), and short-lived ($10-20$\,sec) \citep{2011Sci...331...55D}. Features that may be the coronal counterparts to type-II spicules were seen in Hi-C\,1 data \citep{2014ApJ...784..134R} and IRIS and SDO/AIA data \citep{2017ApJ...845L..18D}, but co-spatial and co-temporal high-resolution observations of the chromosphere, transition region and corona are needed to determine the correlation. 

One of the solar features targeted by Hi-C\,2 is active region moss, which samples the cool footpoints of hot active region loops in this passband  \citep{1994ApJ...422..412P, 1999SoPh..190..409B, 1999SoPh..190..419D, 1999ApJ...520L.135F, 2000ApJ...537..471M, 2003ApJ...590..502D, 2008ApJ...677.1395W}. The Hi-C\,1 data showed short-lived brightenings in the moss which appeared to correlate with AIA $304$\,\AA\ data \citep{2013ApJ...770L...1T}, a possible signature of coronal nanoflares in the transition region footpoints. The cooler passband of Hi-C\,2, along with the co-observations, will allow for studies that correlate the brightenings in the lower corona and transition region with small-scale chromospheric dynamics in AR cores.  

The Hi-C 2 payload was launched on July 27, 2016 (Hi-C~2.0). Unfortunately, an electrical short in the shutter wire prevented the camera shutter from operating and no science data were collected. Some minor changes were made subsequent to the 2016 launch, most notably the repair of the shutter cable and the addition of a Hall effect sensor to verify shutter operation pre-flight. The rocket was successfully re-launched in the summer of 2018, Hi-C\,2.1. In this paper, we present the details of the Hi-C\,2.1 instrument, as well as flight and data performance.  

\section{Experiment Description} \label{sec:inst}

The design and flight of Hi-C\,1 is reviewed in detail in \cite{kobayashi2014}.  For Hi-C\,2.1, there were  two major modifications to the published design.  First, a new coating was applied to shift the passband from 193\,\AA\ to 172\,\AA. Second, the original Hi-C camera was replaced by a new camera capable of significantly lower readout noise.  

In this section, we briefly describe the experiment, including the optical instrument, the camera, and avionics systems.  Much of the experiment remains unchanged from Hi-C\,1, changes from the original instrument configuration are called-out specifically.

\subsection{The Experiment}

\begin{figure*}[h!]
\centerline{\resizebox{6.in}{!}{{\includegraphics{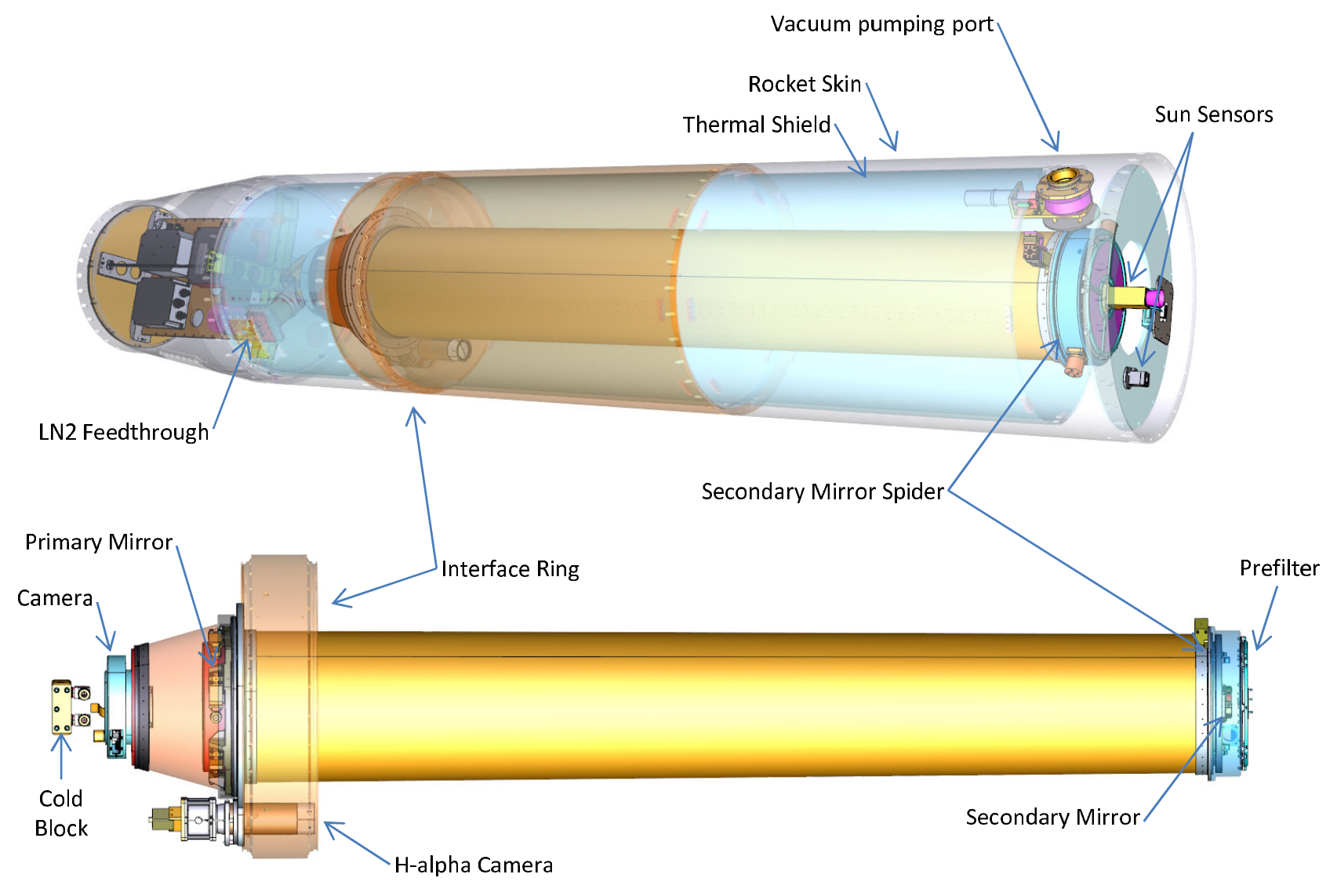}}}}
\caption{The Hi-C experiment layout, indicating locations of the major subsystems \citep{kobayashi2014}.}
\label{experiment}
\end{figure*}

Hi-C is composed of a Ritchey-Chr\'etien EUV telescope, a CCD camera, and a context telescope contained in a standard NASA 22-inch diameter rocket shell. The vehicle layout is shown in Figure~\ref{experiment} with all major components  indicated. The EUV telescope has a plate scale of 0.129~arcsec/pixel and multilayer coatings on the full aperture of the optics. Out-of-band wavelengths are eliminated by front aperture and focal plane filters. The detector for Hi-C\,2.1 is a 2k$\times$2k back-illuminated CCD, providing high quantum efficiency, low noise, and rapid readout for high image cadence. An H-\textalpha\ telescope with NTSC (TV) output is included for real-time pointing verification during the flight.

\subsection{Telescope} \label{sec:tele}

The Hi-C telescope is a classical aplanatic Ritchey-Chr\'etien design using spherical aberration to compensate for field curvature. It consists of a 220\,mm aperture $f/9.1$ primary mirror and a 30\,mm secondary mirror with an inter-optic spacing of 1830\,mm, yielding an effective focal length of 23.9\,m and providing 0.129\arcsec\ per 15\,\textmu m pixel. The RMS spot radius of the optical design is under 2.5\,\textmu m over the entire 4.4\arcmin $\times$ 4.4\arcmin\ field of view (FOV). Taking into account the surface figure and roughness with the mirrors figured to the tolerances listed in Table~\ref{param}, the telescope achieves 0.25\arcsec\ resolution.  The surface figure was measured to 0.5\,nm (RMS) accuracy during manufacturing to determine both mid- and high-frequency components of the mirror roughness. Performance of the telescope was initially verified against a calibrated optical flat; performance on Hi-C\,1 flight confirmed that the telescope optical properties meet its required specifications \citep{kobayashi2014}.

\begin{table}[h!]
\begin{center}
\caption{Hi-C\,2.1 Telescope Optical Design.}
\label{param}
\begin{tabular}{l r l r} \hline
{\bf Telescope:} & &\hspace{0.7cm}{\bf Primary Mirror:} & 
\\
\hspace{.1cm}  Focal Length & 23.9 m &\hspace{0.7cm}  Radius of 
Curvature & 
4000$\pm$4.0 mm \\
\hspace{.1cm}  Plate Scale & 114 $\mu$m/arcsec &\hspace{0.7cm}  
Diameter & 240 mm \\
\hspace{.1cm}  Focal ratio & F/109 &\hspace{0.7cm}  RMS slope error 
& 
0.4 \textmu rad \\
\hspace{.1cm}  Field of View & 4.4$\times$4.4 arcmin &\hspace{0.7cm} 
\\
\hspace{.1cm}  RMS Spot Diam. & 0.07 arcsec &\hspace{0.7cm}{\bf 
Secondary Mirror:} &\\
\hspace{.1cm}    (F.O.V. averaged) & &\hspace{0.7cm}  Radius of 
Curvature & 370.9$\pm$0.5 mm \\
{\bf CCD Camera:} & &\hspace{0.7cm}  Conic & -1.14$\pm$0.10 \\
\hspace{.1cm}  Sensor Size & 942.5 mm$^2$&\hspace{0.7cm}  Diameter & 30 mm \\
\hspace{.1cm}  Plate Scale & 0.129 arcsec/pixel &\hspace{0.7cm}  RMS slope 
error & 
0.1 \textmu rad \\ \hline
\vspace{-0.5in}
\end{tabular}
\end{center}
\end{table}

For the details on the original telescope mount, focus, and alignment process, performed prior to Hi-C\,1 and~2 flights, see \cite{kobayashi2014}.  Hi-C~2 landed at about twice the normal velocity.  In order to ensure that the telescope performance was unchanged, Hi-C~2 was returned to the Smithsonian Astrophysical Observatory (SAO) to confirm that it remained in alignment.  The Hi-C telescope was placed on the SAO alignment bench, and the original telescope alignment procedure was re-run using a Zygo interferometer and a NIST-calibrated reference flat.  There was a small added misalignment between the reference cube and the telescope line of sight when compared to the pre-flight alignment, but the focal plane and the imaging performance were unchanged. The slight misalignment resulted in a measurable added astigmatism, but the low spatial frequency of this defect resulted in an acceptable slope error, and thus an unchanged point spread function of the optical system.

\subsection{Camera} \label{sec:cam}

The camera for Hi-C~2 was custom built at MSFC for this Sounding Rocket. The primary driver for camera replacement was to reduce the noise in the images while maintaining a high-cadence. The Hi-C\,1 camera noise level was between 77 and 102 $e^-$rms \citep{kobayashi2014}. The new Hi-C~2 camera noise levels are about an order of magnitude less, between 9 and 13 $e^-$rms.  

The science camera uses a CCD230-42 back-illuminated, astro-processed sensor manufactured by Teledyne e2v that is operated in full-frame mode. This sensor has $2048 \times 2048$ $15~\mu$m square light-collecting pixels with $50$ non-active pixels in the image read-out registers, and is operated in a readout mode that includes two overscan pixel columns. The non-active and overscan pixels are used for image calibration. The sensor utilizes four read-out registers, or taps, simultaneously with a total read-out time of $2.35$~sec for a full-frame of $2152 \times 2064$ pixels. The pixels are digitized with a 16-bit resolution and the resulting images are sent to the onboard computer via spacewire. 

The sensor is operated in non-inverted-mode (NIMO) to increase the read-out speed and maintain a high-cadence of science data, though it results in higher dark current at a given temperature compared to standard inverted mode (IMO). The CCD is cooled via liquid nitrogen (LN2) to maintain a temperature below $-65^\circ$C during observations to minimize the higher dark current. LN2 is pumped into the payload and through a copper cold block that is connected to the copper CCD holder via a cold strap. The cold block rose from $-122.2^\circ$C to $-117.0^\circ$C during observations. The cold block is significantly colder than the CCD, the CCD temperature dropped from $-78.4^\circ$C to $-79.1^\circ$C during observations. During lab testing we determined that the dark current reached a floor of $\sim 3$~electrons/sec when the sensor was below $-65^\circ$C; the CCD was below this floor during all of the data acquisition during flight. 

Camera characterization in the laboratory included: dark current as a function of temperature, read noise, bias, linearity and saturation, and gain measurements. Each tap has slightly different characteristics, which are described in Section~\ref{sec:data} and summarized in Table~\ref{tab:data_info} therein. 

\subsection{Electronics} \label{sec:elec}

Hi-C's avionics consist of a Data Acquisition and Control System (DACS), low noise power supply, shutter drive, signal conditioning vacuum valve controller, and two vacuum gauges.

The DACS controls camera operations and performs data collection, processing, and transmission; it is an x86 architecture computer running Linux Fedora 19.  The hardware consists of a CompactPCI backplane, a conduction-cooled Single-Board Computer, SBC, (AiTech C802), Spacewire card, digital I/O board, and 500 GByte solid-state storage hard disk drive, and a power supply; these components are packaged in a custom-designed chassis.  The digital I/O board, custom-designed at MSFC, contains a 16-bit parallel interface that is used to send formatted image data to the 10 Mbps telemetry system. The Spacewire card interfaces to the 
science camera.
 
The Hi-C DACS has 3 interfaces to the telemetry system. The 16-bit parallel interface on the digital I/O board, custom designed at MSFC, is used for downlink of acquired science images. A serial uplink channel is used for command uplink, and a serial downlink channel is used for digital housekeeping and status data. Additional telemetry channels are used for analog housekeeping data (temperatures, power supply voltages, supply currents, and camera shutter drive current) via the signal conditioning, and for the video signal of the H-\textalpha\ camera.

While uplink commands are available, they are only used for redundancy; the DACS flight software is designed to operate without need for uplink commands.  Instead, timer commands from the rocket's timer system are used to initiate changes in operating mode: standby, dark frame acquisition, observation, and shutdown.  Additional uplink commands allow for exposure time changes, in case the downlinked images show saturation or severe under-exposure. No uplink commands were sent during the Hi-C\,2.1 flight. 
 
The DACS and flight camera are capable of sustained 4.4\,s cadence with a 2.0\,s exposure time.  The full-resolution images are saved as individual files in standard FITS format on a solid-state disk.  The downlinked images used for pointing were resampled to a 1026 x 1032 resolution and downlinked for real-time display. These downlinked images were used to confirm instrument pointing and assess exposure time during flight. The downlinked images used for science backup data were full 2152~x~2064 pixel resolution in the event that the instrument was un-recoverable after flight.
 
A requirement unique to Hi-C was to monitor the internal vacuum and operate the vacuum valve during the flight, as well as during the countdown phase before the DACS is powered on. The vacuum valve is opened just before the main telescope door, to equalize pressure and minimize the chance of the thin front aperture filter being damaged by a sudden rush of air due to a pressure imbalance when the main telescope door is opened. The DACS is turned on only 10 minutes before flight with the rest of the telemetry system. Separate power is provided via a dedicated payload umbilical and used to operate the vacuum valve, monitor instrument vacuum, and cool the camera CCD to flight operating temps without needing the DACS or telemetry subsystems. This requirement was addressed by using RS-485 interfaces that can be put into a tri-state mode; both the DACS and the Ground Support Equipment (GSE) computer can take control of the vacuum valve controller and vacuum gauge. External GSE is used to control cooling of the science camera CCD.

The low noise power supply provides 8 secondary voltages isolated from the primary 28 Volt input for the science camera and supporting electronics. The power supply has low ripple and switching noise to enable the low read noise requirements for the science camera. The power supply also provides opto-isolated monitor outputs to telemetry for voltages and currents.

\subsection{Multilayer Coatings}

\begin{figure*}[h!]
\begin{center}
\resizebox{0.49\textwidth}{!}{\includegraphics[angle=0,scale=1.]{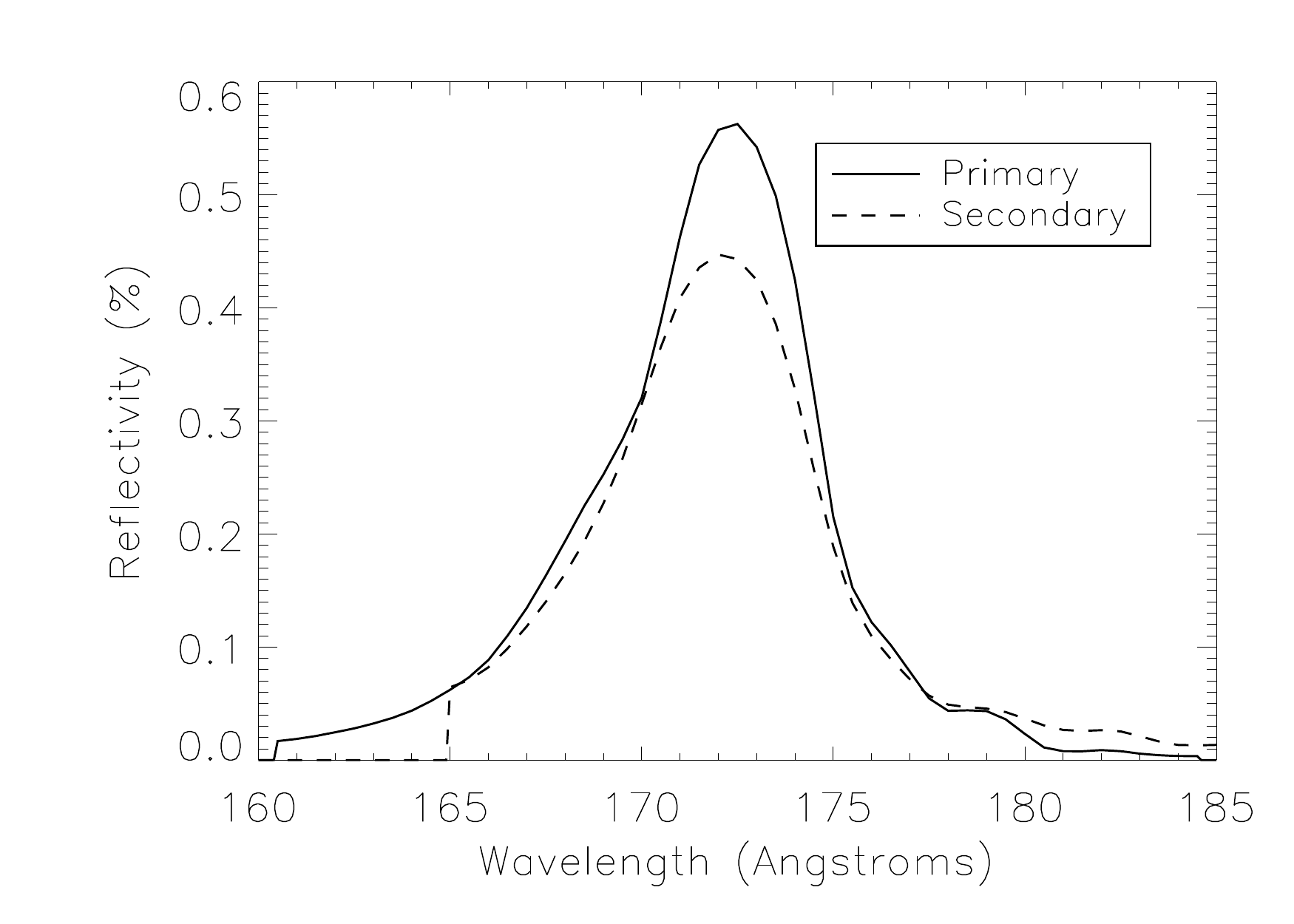}}
\resizebox{0.49\textwidth}{!}{\includegraphics[angle=0,scale=1.]{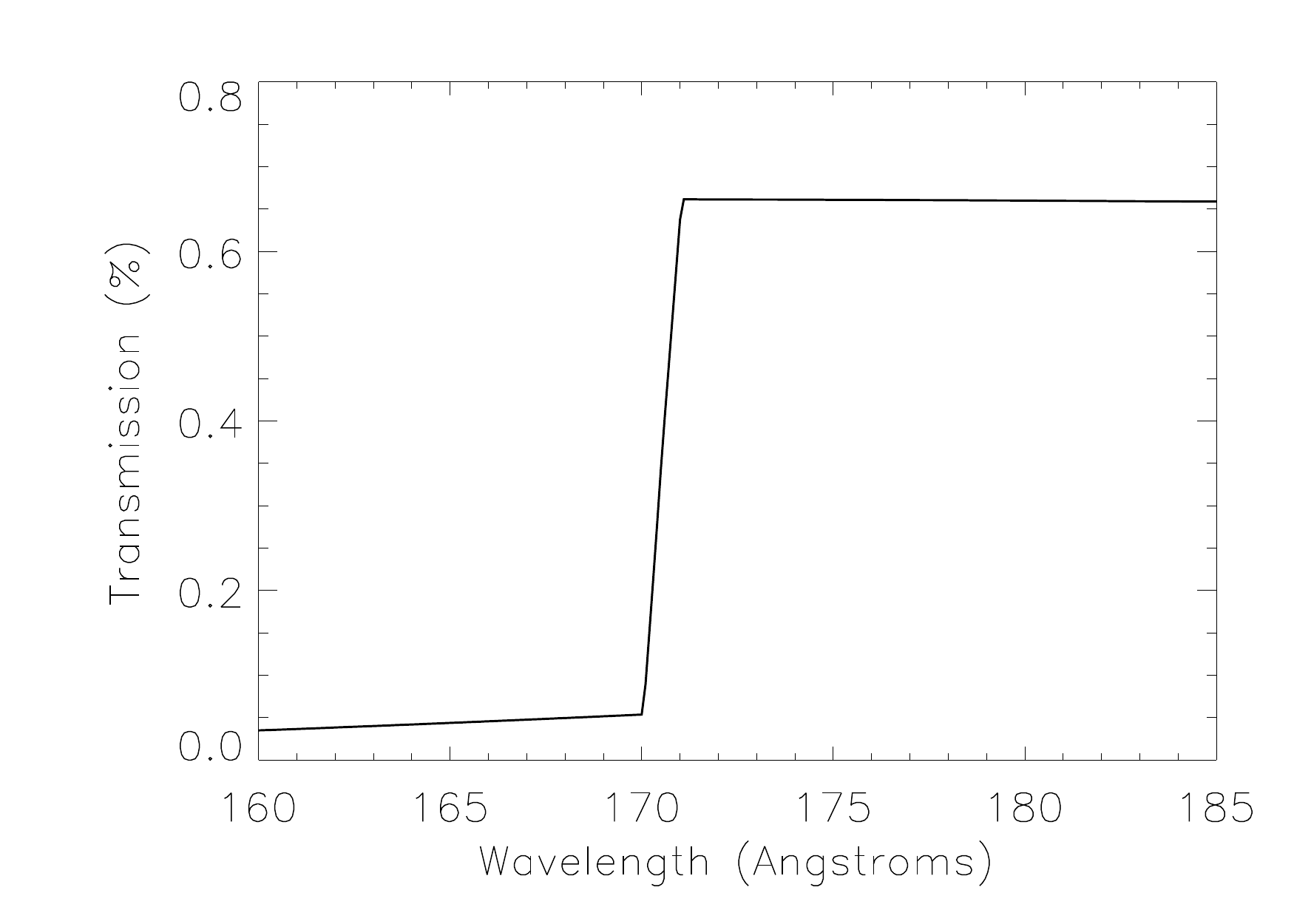}}
\caption{Left: The reflectance of the multilayer coating measured from a witness sample of the primary (solid) and secondary (dashed).  Right: Theoretical transmission of the individual entrance and focal plane filters.}
\label{fig:reflect}
\end{center}
\end{figure*}

The Hi-C\,1 primary mirror was removed from the telescope, cleaned, and re-used for Hi-C~2. The secondary mirror was new for Hi-C~2. The mirrors were coated with periodic Al/Zr multilayers designed for high reflectance near normal incidence over a narrow spectral band centered near 17.2~nm wavelength. Multilayer interference coatings comprising Al and Zr layers have been demonstrated in recent years to provide high reflectance at wavelengths longer than the Al L-edge near 17\,nm, where photoabsorption in the Al layers is relatively low \citep{voronov2011}.  Al/Zr multilayer coatings also have been shown to have good stability and low stress. For the aluminum layers, an Al-Si alloy containing 1\% Si (by weight) can be used in place of pure Al in order to achieve smoother Al-Zr interfaces and thus higher reflectance \citep{zhong2012,windt2015}.   The coatings designed for Hi-C~2 thus contain N=20 repetitions of Al.99Si.01-Zr bilayers of thickness d=8.75 nm (dAl-Si=5.25 nm and dZr=3.5 nm). The coatings were deposited by magnetron sputtering, using a deposition system described in \citet{windt1994}.  Sputter target purities were 99.95\% for Zr and 99.999\% for Al.99Si.01. DC power supplies were used in constant-power mode (400W) for both materials. The deposition system achieved a base pressure of $1.5 \times 10^{-7}$~Torr after pump-down, and Ar (99.999\% purity) was used as the working gas, fixed during deposition at a pressure of 1.6 mTorr using a closed-loop gas-flow system. Si witness samples were coated along with the telescope mirrors. The EUV reflectance of the coated mirrors was measured as a function of wavelength, at $2^\circ$ incidence, using a reflectometer system with a laser-produced-plasma source, also described in \citet{windt1994}. The measured reflectance as a function of wavelength of a witness sample is shown in Figure~\ref{fig:reflect}, showing peak reflectance of ~55\% and a spectral band-width of 0.4 nm FWHM. 

\subsection{Filters}

The Hi-C experiment included two thin-film filters, one at the entrance to the telescope and one in the focal plane.  Both filters were 150\,nm of aluminum film mounting on 5 lines per inch nickel mesh in custom designed frames made by Luxel.  The material is identical to the entrance and focal plane filters of the Solar Dynamics Observatory Atmospheric Imaging Assembly (SDO/AIA) 171\,\AA\ channel \citep{lemen2012}, but with much finer mesh.  Transmission curves for the filters are shown in the right hand side of Figure~\ref{fig:reflect}. When calculating the transmission of the filters, we assume that the mesh has 98\% area.  

As with Hi-C\,1, the focal plane filter was situated close enough to the telescope focus to create a mesh shadow-pattern on the images. The removal of this grid  pattern in post-processing is described in Section~\ref{sec:data}.

\subsection{H-\textalpha~Camera and Pointing GUI}\label{sub:halpha}

The primary function of the H-\textalpha\ camera system is to confirm that the telescope is pointing at the selected target and to provide context for gross pointing maneuvers in case of significant pointing error. (The EUV images from the science camera are used for fine pointing correction.) The H-\textalpha\ camera is composed of a 712\,mm effective focal length negative telephoto lens system, a Day-Star Corporation H-\textalpha\ narrow band-pass (0.6\,\AA) filter centered at 6563.28\,\AA, neutral density filters, and a Sony monochrome camera.  This system is identical to the one flown on the Tunable XUV Image (TXI) rocket, which produced a full Sun image with good contrast and resolution, allowing us to recognize sunspots, and confirm pointing orientation.   

\begin{figure*}[h!]
\begin{center}
\resizebox{0.49\textwidth}{!}{\includegraphics[angle=0,scale=1.]{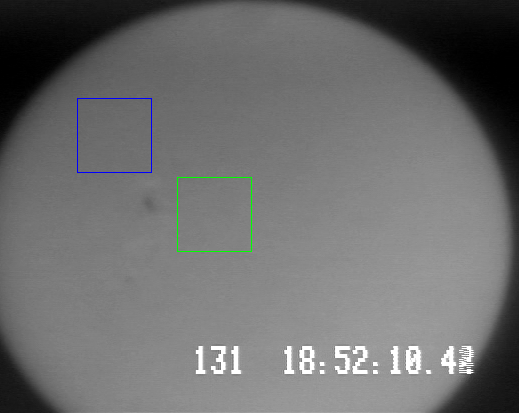}}
\caption{Image from the real-time H-\textalpha\ pointing GUI during on-ground testing. The blue box shows the intended target and the green box shows the current target.}
\label{fig:halphapointing}
\end{center}
\end{figure*}

To improve the functionality of the H-\textalpha\ camera system for Hi-C\,2.1, a graphical user interface (GUI) application was developed to use the H-\textalpha\ images to determine, in real-time, the pointing of the telescope relative to the pre-determined target.  An image processing algorithm was developed by the University of Central Lancashire (UCLAN) to locate the Sun center within the H-\textalpha\ image, which was then cropped to the edge of the determined solar limb. Based upon where the Sun center is located in this image and the offset between the H-\textalpha\ camera and science camera FOV (measured pre-flight), a box was outlined in real-time on the H-\textalpha\ image to display the current pointing of the EUV camera. A second box was also displayed for the pre-planned FOV in order to determine any differences in telescope pointing (Figure~\ref{fig:halphapointing}). Thus, this information allowed for any significant pointing differences to be recognized immediately and a decision made on any need for subsequent adjustment. During the flight, H-\textalpha\ data were captured at a rate of one in every ten frames and saved as FITS files for further post-flight analysis. A similar application is being developed by UCLAN to aid in pointing of the Marshall Grazing Incidence X-ray Spectrometer (MaGIXS) instrument using data from the MaGIXS slit jaw camera. 

\subsection{Radiometry and Exposure Time Estimate} \label{sec:rad}

\begin{figure*}[h!]
\begin{center}
\resizebox{0.49\textwidth}{!}{\includegraphics[angle=0,scale=1.]{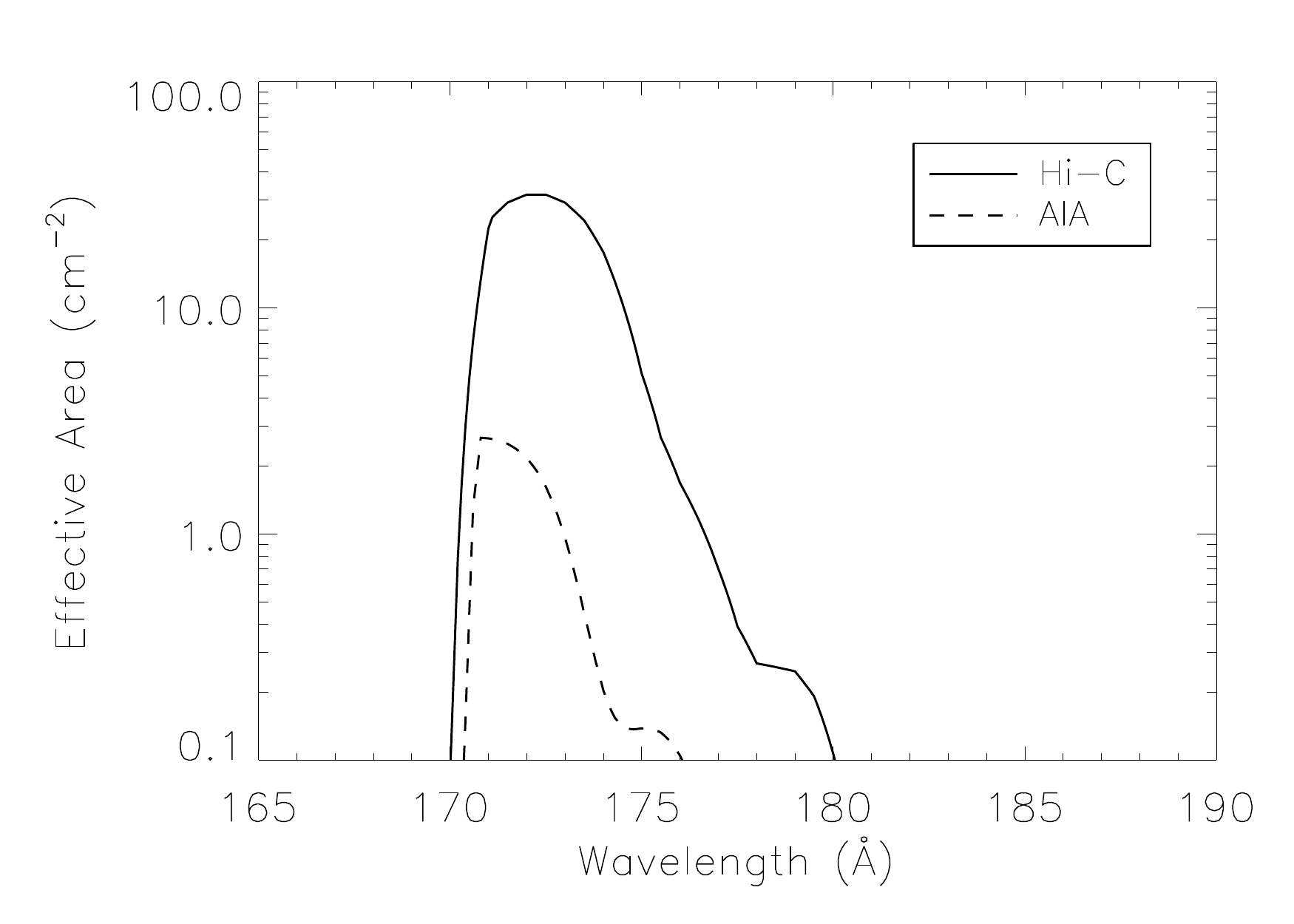}}
\resizebox{0.49\textwidth}{!}{\includegraphics[angle=0,scale=1.]{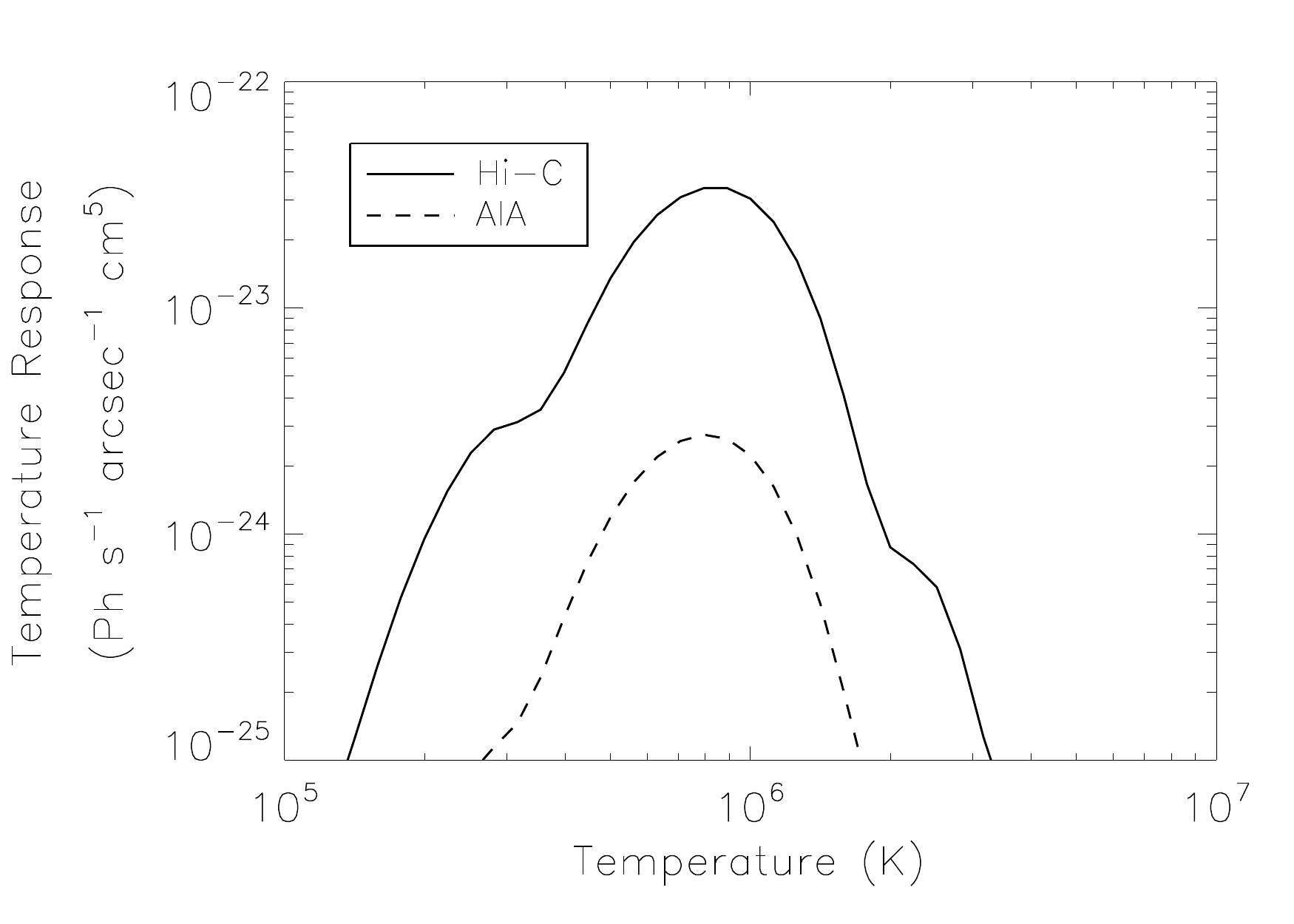}}
\caption{ The effective area (left panel) and temperature response (right panel) of the Hi-C\,2.1~172\,\AA\ passband (solid) and AIA 171\,\AA\ channel (dashed).}
\label{ea_tr}
\end{center}
\end{figure*}

Before flight, it was important to calculate the expected throughput of the instrument and predict an adequate exposure time to prevent saturation or under-exposure.  Using the geometric area of the telescope, the measured reflectance curves of the multilayers (Figure~\ref{fig:reflect}, left panel), transmission of the entrance and focal plane filters (Figure~\ref{fig:reflect}, right panel), and the expected quantum efficiency of the CCD, we have calculated the effective area of  Hi-C~2~172\,\AA\ passband; this is shown in Figure~\ref{ea_tr}. For comparison, we show the effective area of the AIA 171\,\AA\ channel with a dashed line, which was calculated using the options of EVE normalization and the time-dependent correction for the Hi-C\,2.1 launch date.  The effective area of the Hi-C\,2.1~172\,\AA\ passband is broader in shape, shifted in wavelength (peak is at 170.8\,\AA\ in AIA and at 172\,\AA\ in Hi-C\,2.1), and 11.9 times larger than the AIA\,171\,\AA\ channel. Using these two effective area curves, we calculate the temperature response
for both Hi-C and AIA assuming standard Chianti ionization equilibrium and coronal abundances from \cite{schmelz2012}.  These curves are shown on the right side of Figure~\ref{ea_tr}.  The Hi-C\,2.1~172\,\AA\ and AIA\,171\,\AA\ temperature responses both peak at Log\,T=5.9.  

Hi-C\,2.1 was expected to transmit 10.7 times more signal per arcsec than the AIA 171\,\AA\ channel, but it has roughly 22 pixels for each AIA pixel.  Based on this calculation and our previous experience with the first flight of Hi-C \citep{winebarger2014a}, we estimated that a 2\,s exposure time was appropriate for this flight.  
 
\begin{figure*}[h!]
\begin{center}
\resizebox{0.9\textwidth}{!}{\includegraphics[angle=0,scale=1.]{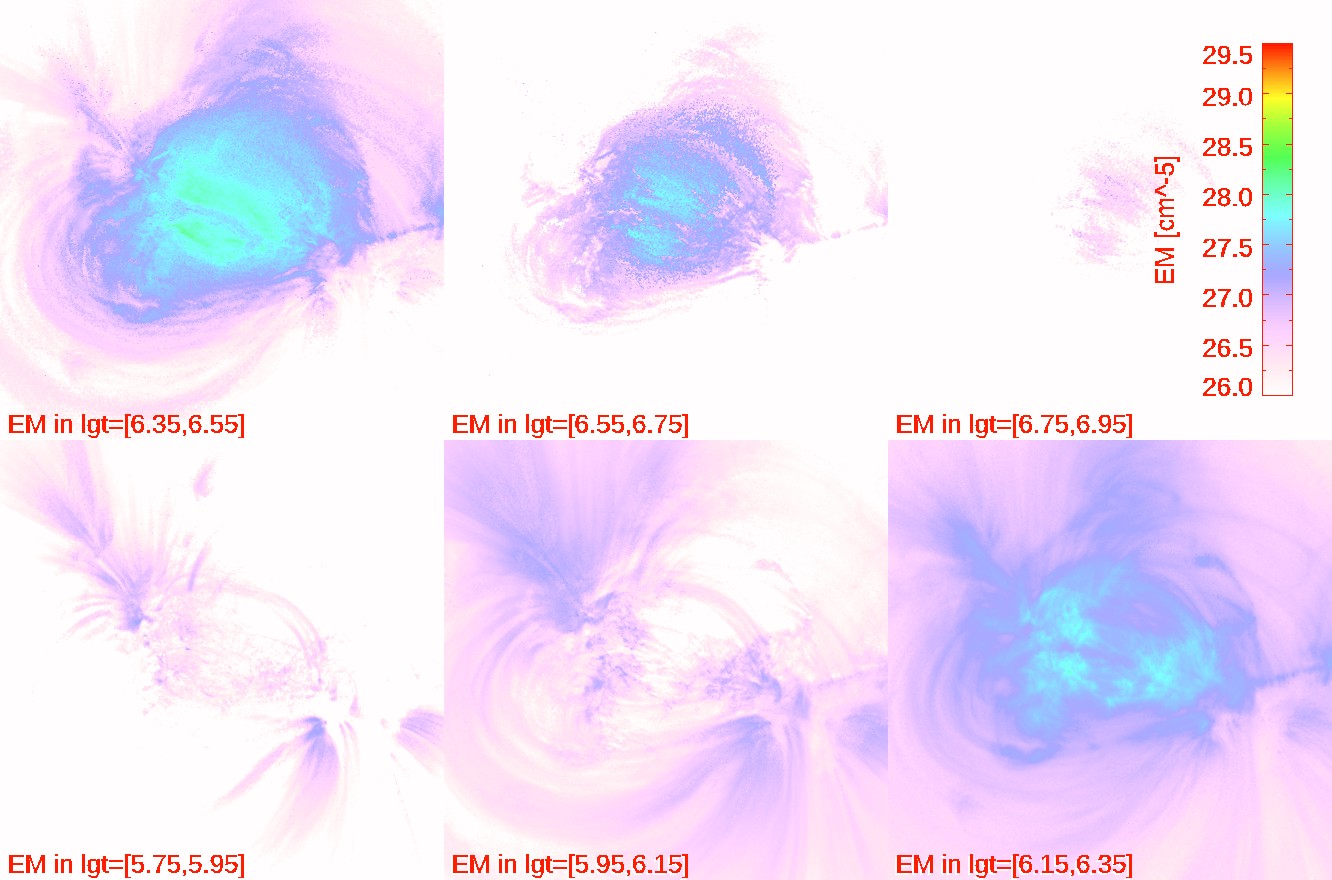}}
\caption{The emission measure distribution calculated from co-temporal and co-spatial AIA data.}
\label{fig:em}
\end{center}
\end{figure*}

After flight, we used co-temporal AIA data to validate the radiometry of Hi-C\,2.1.  Because the shape of the Hi-C\,2.1 passband was slightly different than the AIA 171\,\AA\ channel, we use the six AIA EUV channels taken closest to the time that Hi-C\,2.1 was at the peak of its flight (where Hi-C\,2.1 data suffers least from atmospheric absorption, see Figure~\ref{fig:absorb}) and calculate an emission measure curve at every AIA pixel using the method of \cite{cheung2015}, this emission measure distribution is shown in Figure~\ref{fig:em}.  We then convolve the emission measure with the Hi-C\,2.1 temperature response function and predict the expected counts in Hi-C\,2.1.  We find that the counts in Hi-C are 15\% larger than expected.  We assume this 15\% is within the uncertainty of the AIA calibration and the Hi-C component level effective area calculation.

\section{Flight Performance} \label{sec:flt}
\begin{figure}[ht]
\begin{center}
\includegraphics[width=0.7\textwidth]{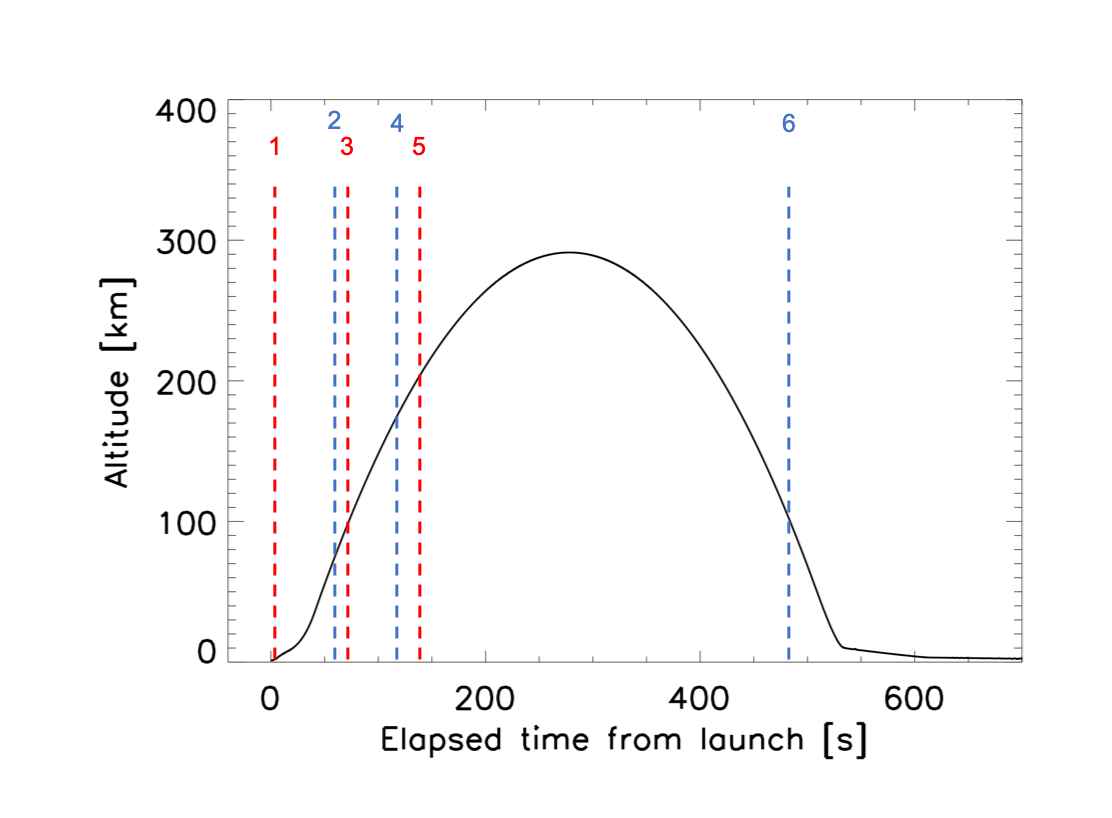}
\caption{The altitude of the Hi-C\,2.1 rocket determined from White Sands Missile Range radar data as a function of elapsed time from launch (18:54:00 UT).  The event times listed in Table~\ref{tab:timeline} are labeled.}
\label{fig:timeline}
\end{center}
\end{figure}

Hi-C\,2.1 was launched at 18:54~UT on May 29, 2018 from White Sands Missile Range.  The target of observation was Active Region 12712. The Solar Pointing and Aerobee Control System (SPARCS) maintained a constant target for the duration of the flight.  For 335\,s, Hi-C\,2.1 recorded full detector ($\sim$2k$\times$2k) images with a 2\,s exposure at a cadence of 4.4\,s.  Because the time on the Hi-C onboard DACS drifts, an adjustment of 126\,s was applied to all data headers in post-processing. Table~\ref{tab:timeline} provides the timeline of the Hi-C\,2.1 rocket flight. Figure~\ref{fig:timeline} provides the height of the sounding rocket as a function of time, determined from White Sands Missile Range radar measurements.  The events given in Table~\ref{tab:timeline} and the approximate height at which they occurred are indicated in this figure.


\begin{center}
\begin{table}[ht]
\caption{Hi-C\,2.1 Flight Event Timeline (May 29, 2018)}
\begin{tabular}{lll}\hline
{\bf} & {\bf Event} & {\bf Time (UTC)}\\ \hline
0 & Launch        &    18:54:00.41 \\
1 & Start Dark Exposures  &  18:54:04\\
2 & End Dark  Exposures  &  18:55:06\\
3 & Shutter Door Open     &   18:55:12 \\
4 & Fine Pointing    &    \\
 & [Ring Laser Gyroscope (RLG) Enable] & 18:56:01\\
5 & Data Acquisition     &     18:56:21\\
6 & Shutter door close    &   19:02:00 \\ \hline
\end{tabular}
\label{tab:timeline}
\end{table}
\end{center}

\begin{figure*}[h!]
\begin{center}
\includegraphics[width=0.7\textwidth]{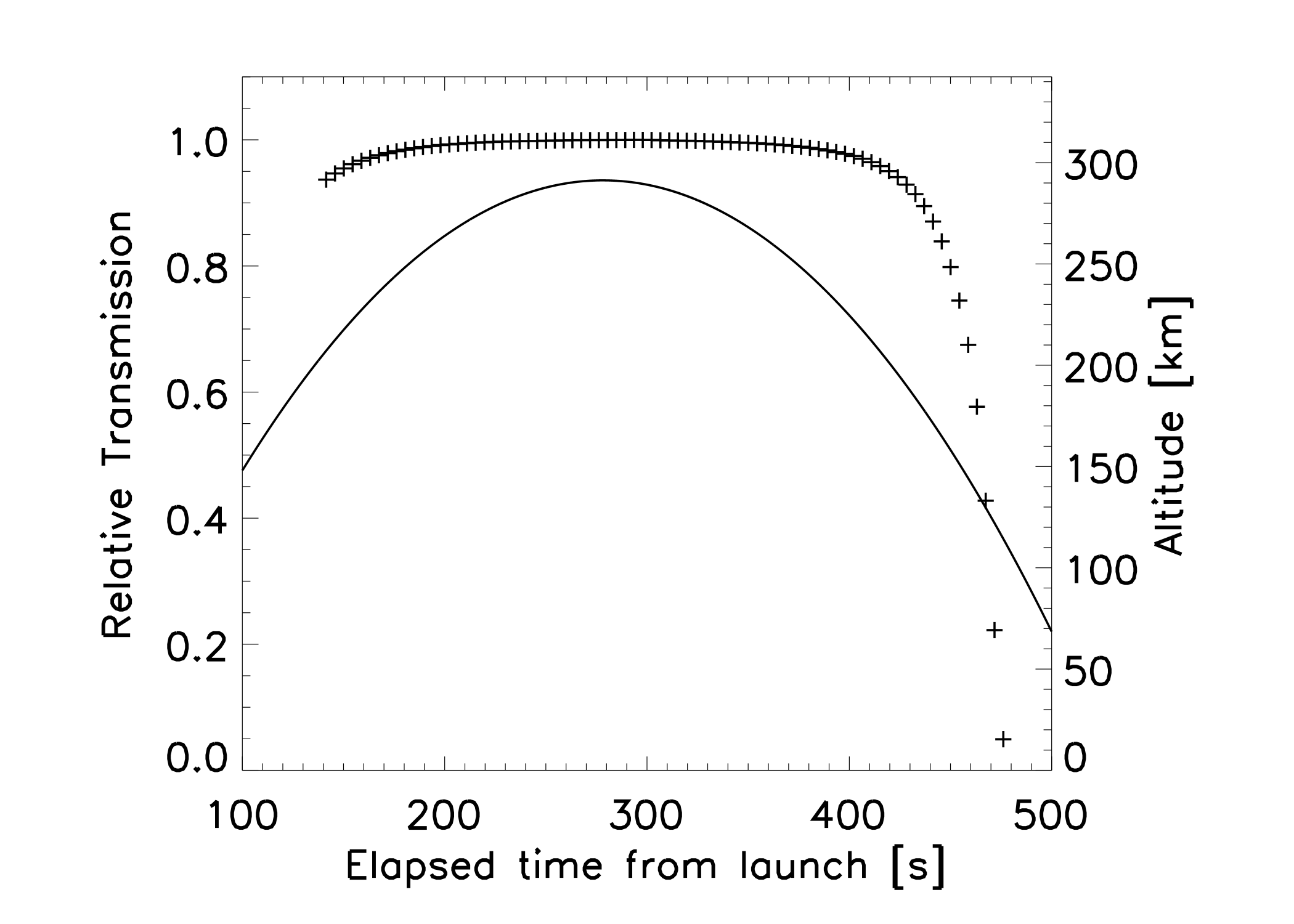}
\caption{The relative transmission of the Hi-C\,2.1 data (plus symbols) and payload altitude (solid line) as a function of time elapsed since launch (18:54:00 UT).}
\label{fig:absorb}
\end{center}
\end{figure*}

We use the normalized total intensities of the Level~1.0 processed flight data (processing levels described in Section~\ref{sec:data}) to assess the relative atmospheric absorption of the signal as a function of flight time.  The transmission, shown in Figure~\ref{fig:absorb} in combination with the payload altitude, is calculated as the inverse of the relative absorption (i.e., (absorption coefficient)$^{-1}$). More than 4 minutes of data were unaffected by the atmosphere.  The atmospheric absorption was compensated for in the Level~1.5 processed data set by multiplying the images by their respective absorption coefficient.  These coefficients are provided in the header of this processed set.

\subsection{Pointing} \label{sec:point}

As with the Hi-C\,1 flight, the H-\textalpha\ pointing camera system described in Section~\ref{sub:halpha} did not take on-band images. The wavelength is controlled by a temperature-sensitive etalon, and the heating caused by exposure to full-sunlight under vacuum caused a wavelength shift. The full-sun continuum images were sufficient to verify coarse pointing with the GUI system described in Section~\ref{sub:halpha}, which was further refined using downlinked the science camera images.

\begin{figure}[h!]
\begin{center}
\includegraphics[width=0.7\textwidth]{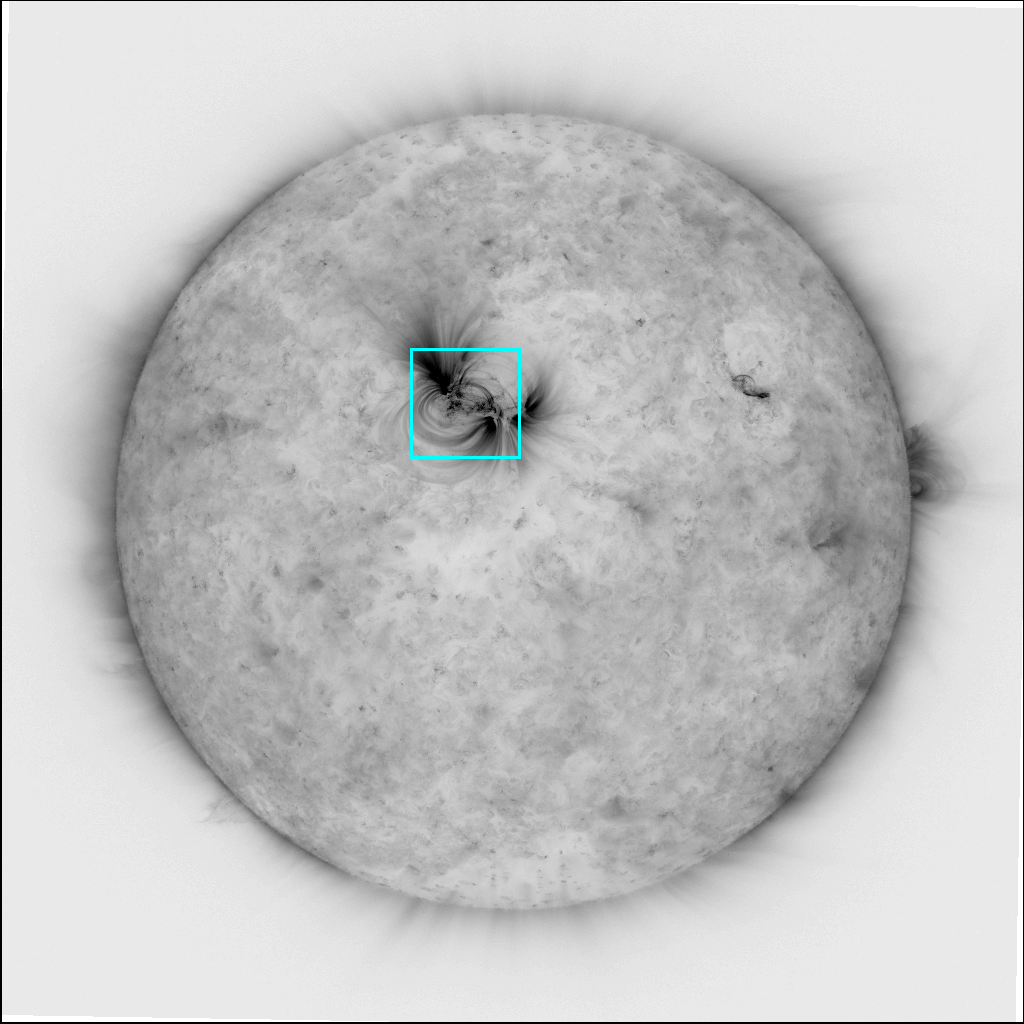}
\caption{The reference AIA 171\,\AA\ data taken at 18:56:57~UT, which was used for determining the absolute pointing. The highlighted box indicates the Hi-C\,2.1 FOV.}
\label{fig:fov}
\end{center}
\end{figure}

To determine the roll offset and absolute pointing post flight, the AIA\,171\,\AA\ image taken at 18:56:57.35~UT was used as a reference against the Hi-C\,2.1 image taken at 18:56:56.64~UT.  The roll offset, found to be $\sim0.985^\circ$ (clockwise about Sun center), is within the tolerances for SPARCS pointing.  Figure~\ref{fig:fov} shows the full-disk AIA\,171\,\AA\ image rotated to this offset. The Hi-C\,2.1 FOV, centered at (-114\arcsec, 259\arcsec), is indicated by the box.  

The full Hi-C\,2.1 image set was co-aligned through a combination of cross-correlation techniques. The target region drifted in the FOV by $<$\,3 pixels due to solar rotation during the flight, as determined by an AIA image set spanning the flight. The headers of the Level~0.5 and higher data files were adjusted to include the best approximation of the absolute pointing for each image, including the fine shifts.

\subsection{Stability and Resolution}
\label{sec:stability_resolution}

\begin{figure}[H]
    \begin{center}
    \includegraphics[width=4in]{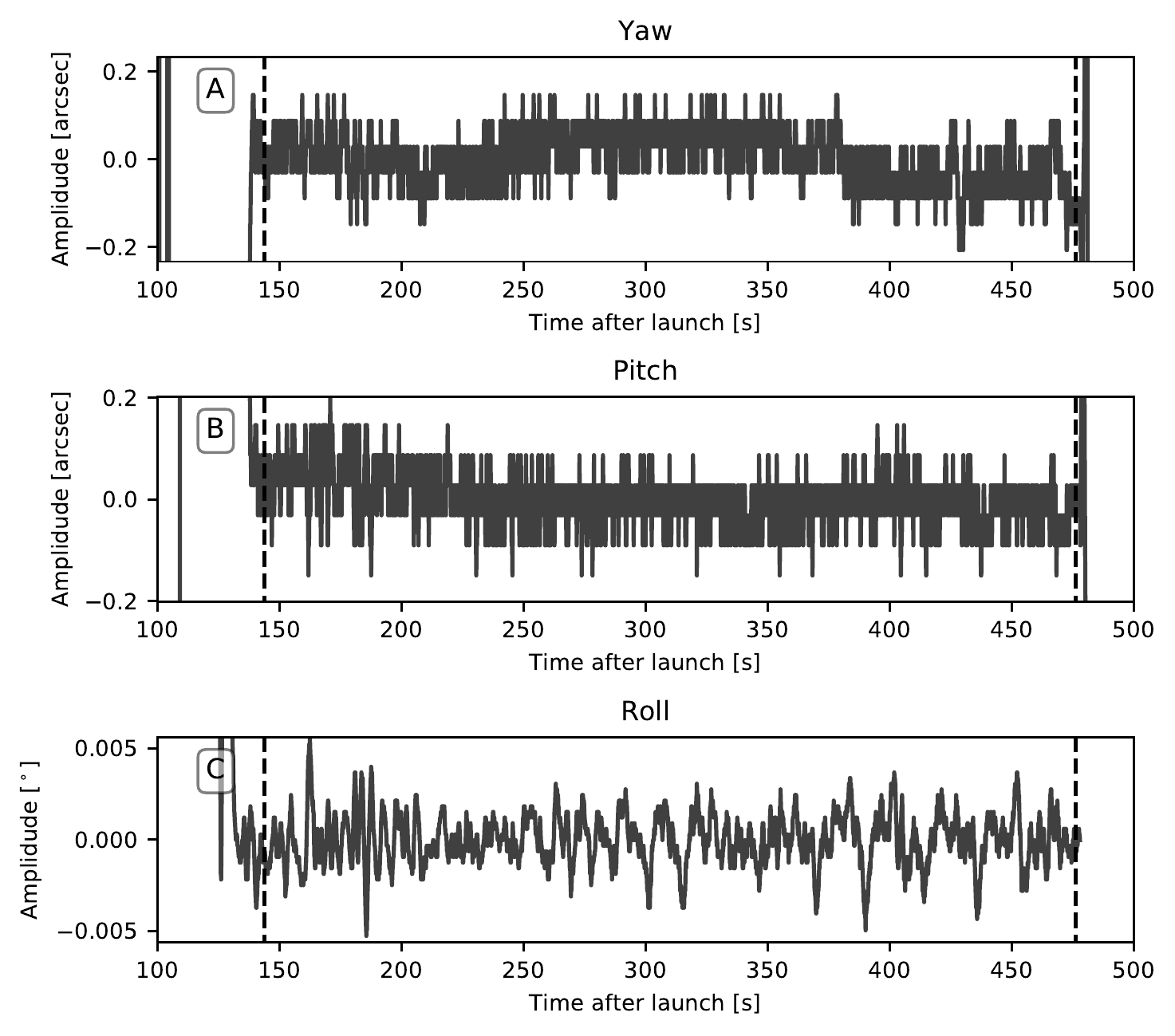}
        \caption{Pointing stability as a function of time during the imaging period. Yaw (A), pitch (B) and roll (C) are plotted for the entire imaging period. Image acquisition period is indicated by dashed black lines. Notice a semi-periodic variation of roll pointing during observation. }
        \label{fig:pointing_and_jitter}
    \end{center}
\end{figure}

Though Hi-C\,2.1 achieved all pointing and stability requirements, enabling the acquisition of the highest resolution 17.2\,nm coronal images ever taken, approximately half of the images captured during flight show signs of greater than expected motion blur. The stability requirements were RMS pitch and yaw jitter of  $<$\,0.3\arcsec\, and RMS roll jitter of $<$\,0.01$^\circ$ for 90\% of the observation time. These requirements were met as shown by recorded jitter as a function of time (Figure~\ref{fig:pointing_and_jitter}). The measured RMS jitter values were $\sigma_{yaw} = 0.06\arcsec$, $\sigma_{pitch} = 0.05\arcsec$  and $\sigma_{roll} = 0.0014^\circ$. Some exposures indicate pointing instability resulting in motion blur and lower spatial resolution. This motion blur is observed to be semi-periodic, impacting every 6-8th frame most severely. The most likely source of motion blur is roll instability, a semi-periodic variation can be seen in measured roll data (Figure~\ref{fig:pointing_and_jitter}C). 

\begin{figure}[H]
    \begin{center}
    \includegraphics[width=\textwidth]{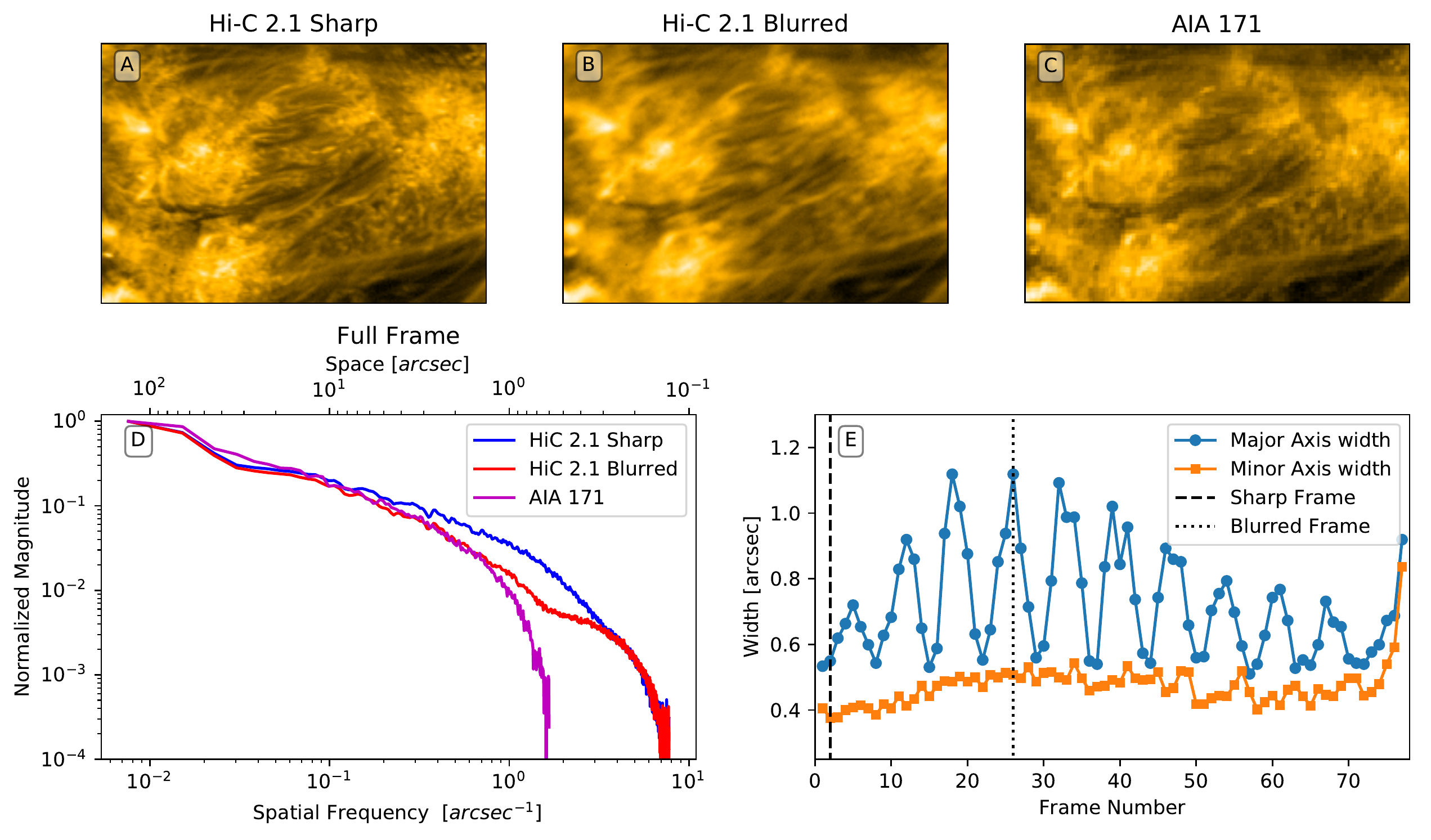}
        \caption{A sharp image (A) and a motion blurred image from the Hi-C 2.1 data set (B) are given with the corresponding AIA 171 sub-image (C).  The full frame averaged FFT (D) plotted with space (top axis) and spatial frequency (bottom axis). The major and minor axis of the FFT width is shown in (E). The vertical lines correspond to the frame numbers of the example sharp (dashed) and blurred (dotted) images. }
        \label{fig:FFT_psf_approx}
    \end{center}
\end{figure}

\begin{figure}[H]
    \begin{center}
    \includegraphics[width=4in]{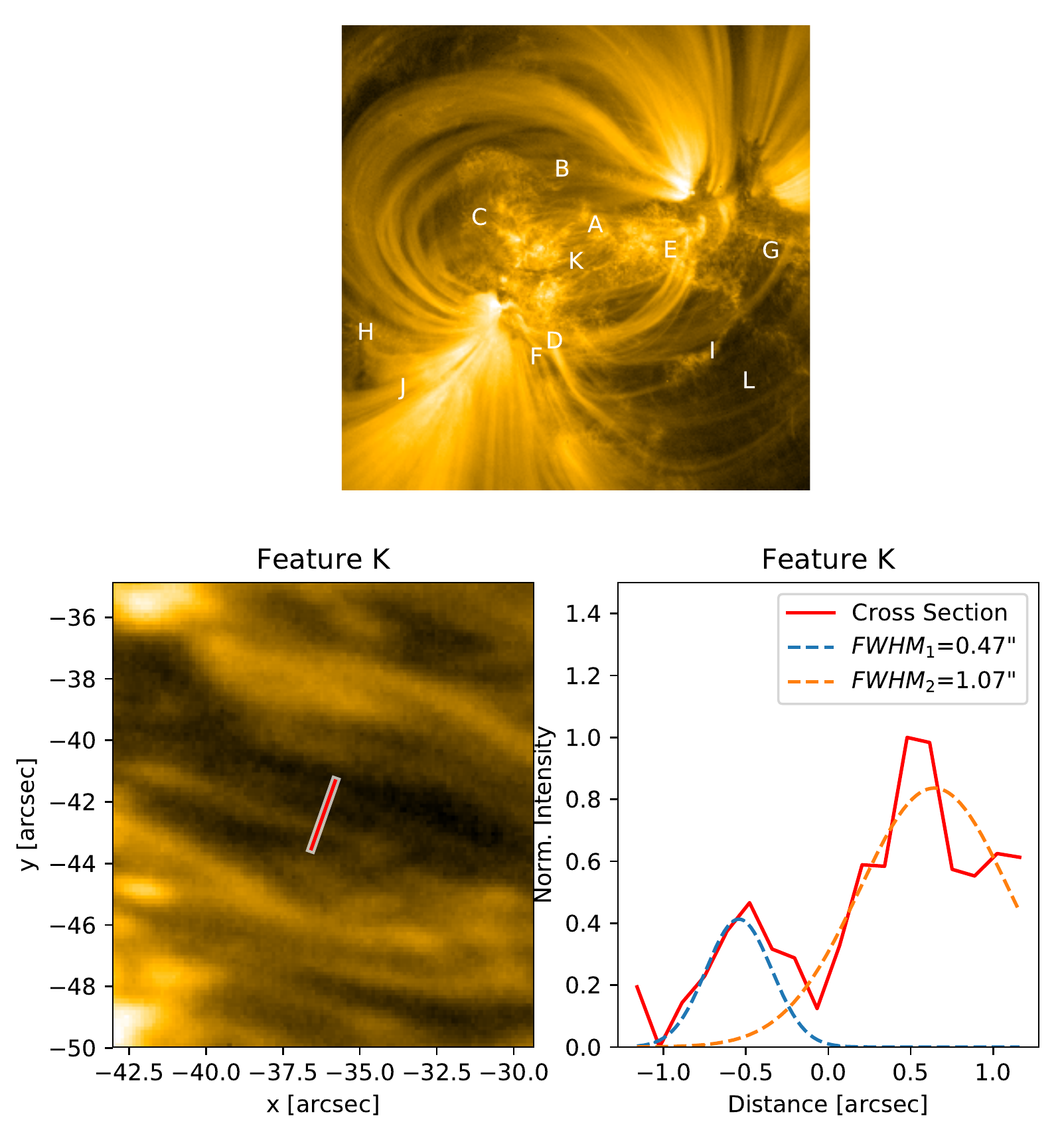}
        \caption{Feature cross sections and fitted widths. Sharp frame of Hi-C\,2.1 data set marked with locations of features sampled (top) which correspond to Table \ref{tab:Feature_widths}. A zoomed-in example of a sample feature (bottom left) with intensity along the cross section and Gaussian fits (bottom right). FWHM of the Gaussian fits are given is given in the legend. }
        \label{fig:example_cross}
    \end{center}
\end{figure}

The image resolution is a function of both the instrument PSF and motion blur. At best, the image resolution must be greater than or equal to 0.26\,\arcsec, twice the pixel plate scale. In Hi-C\,2.1 the resolution changes with time, implying that motion blur was acting to degrade the resolution of some of the images. The spatial resolution has been approximated by Fourier analysis, (Figure~\ref{fig:FFT_psf_approx}), and by analyzing intensity cross sections of fine structures, (Figure~\ref{fig:example_cross}). 

A 2D Fourier transform  describes the spatial frequency information contained in an image \citep{Young2011}. By assessing the spatial frequency content of an image, relative resolution performance of individual frames can be compared and a rough estimate of the image resolution obtained. A similar analysis was applied in \citet{kobayashi2014} for Hi-C\,1 data. In Hi-C\,1, the frequency spectrum followed the slope of AIA 171\,\AA\, frequencies down to at least 3-4~arcsec$^{-1}$ when an average of four sequential Hi-C\,1 frames was used to reduce the noise. Due to time-varying nature of the Hi-C\,2.1 images, no four subsequent frames maintain a consistent resolution, and this method is not as applicable. However, we can still analyze the Fourier information to determine the spatial scale of the noise-floor of a single image.

Figure~\ref{fig:FFT_psf_approx} shows two sub-frames from the Hi-C\,2.1 data set, one that experienced low jitter (A) and one that was significantly blurred (B). The corresponding field of view acquired in the AIA 171\,\AA\, channel is shown (C) for comparison. The azimuthaly averaged full frame frequency power spectrum of the Hi-C\,2.1 frames and AIA 171 \AA\, sub-image are plotted (D). The major and minor axis of width (i.e. extent) of the 2D Fast Fourier Transform (FFT) as a function of time is shown in (E). The widths are approximated by the inflection point where the spectra becomes dominated by a function differing from that which defines the lower frequency components. The FFT width varies between $\approx$0.35-1.1\arcsec\,and oscillates between sharp and blurred with a period of approximately 6-8 images, or 27-36 seconds. 

The curvature of the FFT (Figure~\ref{fig:FFT_psf_approx}D) shows that the blurred image (red line) has a clear inflection point at approximately 0.6\arcsec, indicating that below this level, the image is being impacted by motion blur. The lack of a clear inflection point for the sharp image (blue line) indicates the image is limited by instrument resolution or scale of structures in the image instead of by motion blur. The sharp image FFT deviates from the AIA 171 \AA\, FFT (magenta) at spatial scales of approximately 5\arcsec, while the blurred image FFT (red) deviates from AIA at around 1.5\arcsec. These deviations signify the spatial scales at which Hi-C data are better resolved than AIA. Further, both the sharp and blurred Hi-C curves converge again around 0.3\arcsec, indicating that features below this scale are not well resolved in any single Hi-C\,2.1 image, being limited by shot noise and nearing the instrument Nyquist frequency.

The Fourier spectrum is related to the image resolution, but not exactly equivalent. The spectrum will change based on the spatial frequency content present in the objects being imaged, and so this analysis is dependent on the target selection. The Fourier analysis does clearly show that the Hi-C\,2.1 images have a clear periodic resolution degradation due to motion blur, and gives a limit to spatial scales resolved in the images, but does not provide an unambiguous measurement of image resolution. 

In assessing Hi-C\,2.1 resolution performance, it is also helpful to examine directly the widths of small features in un-blurred frames.  Gaussian fits corresponding to a selection of small features are listed in Table \ref{tab:Feature_widths}. The locations for each feature are noted in Figure \ref{fig:example_cross} along with an example of the cross section and fitting performed. Of these selected features, the smallest resolved Gaussian width is $\approx0.47$\,\arcsec, implying that the image resolution is better than $0.47$\,\arcsec. True resolution can be assessed from the imaged width of a sub-resolution feature such as a point source. However, the features summarized in this list do not represent sub-resolution features, and therefore can only provide an upper bound on resolution. These features were manually selected from a single frame and this analysis does not rule out the possibility of resolved smaller features, especially if they are present in multiple frames.

Combining both FFT and Gaussian width analysis methods, we conclude that the Hi-C\,2.1 resolution is between 0.3 and 0.47 arcsec in images that are not affected by motion blur.

\begin{center}
\begin{table}[ht]
\caption{Feature cross section list. Feature locations are shown in Figure~\ref{fig:example_cross}. Listed location is the pixel index of beginning and end of the cross sectional line, where the first pixel in the image is [0,0]. }
\begin{tabular}{c l c c}
\hline
{\bf Label} & {\bf Location} & {$\mathbf{FWHM_1}$ (\arcsec)} & {$\mathbf{FWHM_2}$ (\arcsec)}\\ \hline
A & [1027, 1133], [847, 963]   &  0.7  & 0.66 \\
B & [887, 989], [602, 719]     &  0.94 &       \\
C & [521, 644], [833, 933]     &  1.61 &       \\
D & [850, 955], [1375, 1520]   &  2.95 &  1.2  \\
E & [1368, 1469], [962, 1075]  &  0.83 &       \\
F & [780, 906],[1436, 1546]    &  2.11 &  0.98 \\
G & [1804, 1919],[978, 1081]   &  0.89 &       \\
H & [19, 127],[1337, 1462]     &  1.07 &  0.90 \\
I & [1559, 1670],[1398, 1520]  &  0.88 &  1.44 \\
J & [206, 310],[1582, 1690]    &  0.83 &       \\
K & [944, 1050],[1008, 1125]   &  0.47  &  1.07  \\
L & [1717, 1834],[1545, 1652]  &  1.1 &       \\\hline
  
\end{tabular}
\label{tab:Feature_widths}
\end{table}
\end{center}

\section{Data} \label{sec:data}

The Hi-C\,2.1 images are processed by subtracting the bias readout pedestal and the dark current, flat-fielding, and correcting for bad pixels.

{\bf Bias:}  The non-active regions of the images (see Section~\ref{sec:cam}) are used to determine the bias pedastal, which varies slightly between quadrants and as a function of time. The bias is calculated and removed per frame and per quadrant. 

\begin{figure}[h!]
\begin{center}
\includegraphics[width=1.0\textwidth]{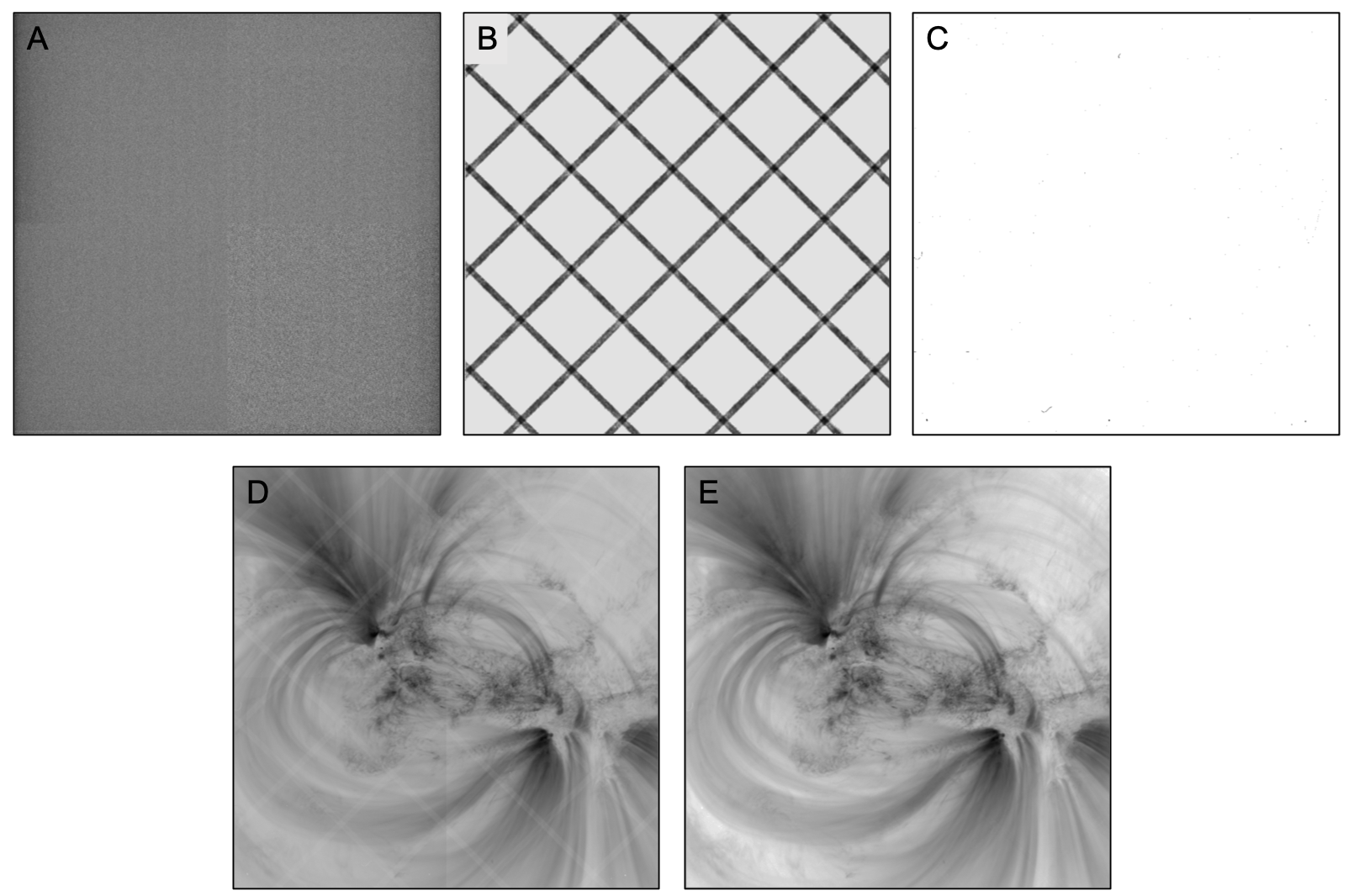}
\caption{The upper panels show the A) master dark, B) master flat, and C) and bad-pixel map respectively.  The bottom panel shows an example of D) the raw Hi-C data (rotated and cropped) and E) a calibrated image. Figures D and E are plotted with an inverse colortable.}
\label{fig:process}
\end{center}
\end{figure}

{\bf Dark Current:} During ascent, 15 dark frames were obtained matching the exposure time of the science images (i.e., the images exposed to sunlight). A master dark frame was created from a median of these images (to remove the contribution of particle hits), after removal of the bias in each of the individual dark frames.  This master dark (Figure~\ref{fig:process}.A) is also subtracted from all of the science images.  

{\bf Flat-field:} The data were further affected by the shadow of the mesh from the focal plane filter, reducing the intensity behind the mesh by up to $\sim$35\%. The mesh pattern and its associated transparency was derived from the blurred, exposed images taken while slewing during the initial fine pointing procedures. A master flat-field image was created to compensate for the mesh, which does not affect the unobscured pixels. The data were divided by this master flat (Figure~\ref{fig:process}.B) to reduce the presence of the grid. 

{\bf Bad pixels:} Finally, a bad pixel map (including pixels obscured by dust, Figure~\ref{fig:process}.C) was generated by applying a threshold to the dark frames and the science images that were blurred during initial fine pointing procedures. The adjusted intensity in the affected pixels is interpolated from the nearest pixels that are not identified as containing dust or bad pixels.  

Raw and calibrated images are shown in Figure~\ref{fig:process} D and E.  (Note that the raw image has been rotated by 90$^\circ$ counter-clockwise prior to processing, in order to place solar north toward the top of the frame, and the non-active pixels have been cropped.)

{\bf Gain:} The camera gain was determined pre-flight (using an Fe55 source) to be 2.5 electrons (e$^{-}$) per data number (DN), with variability less than 0.02 elec\,DN$^{-1}$ between quadrants.

{\bf Noise:} The camera was cooled below the required -65$^\circ$C during flight and therefore achieved low dark current. The remaining noise is dominated by the camera read-noise. The noise in Table~\ref{tab:data_info} is calculated as the median standard deviation per quadrant of the set of bias-subtracted darks taken during the ascent phase of the flight.  The results per quadrant of this set range from $\sim$3.6 to 5.5\,DN (i.e., $\sim$9$-$13.8\,e$^-$.)

The noise of the master dark (created from a median of these images as described above) is $\sim$1\,DN per
quadrant, therefore not adding any significant noise to the science images during processing (described below). \\ 

A summary of the flight data parameters, as described in the preceding sections, is provided below in Table~\ref{tab:data_info}. 

\begin{center}
\begin{table}[H]
\caption{Hi-C\,2.1 Flight Data Summary (*Solar coordinates; solar north at top of frame.)}
\begin{tabular}{ll | l l}\hline
Channel &   172\,\AA\  & Image Size  & 2064$\times$2048\\
Launch Date & May 29, 2018 & Field of View  & 4.44\arcmin$\times$4.40\arcmin \\
Data Acquisition Time & 18:56:22 - 19:01:57 & Pointing  & (-114\arcsec, 259\arcsec)  \\
Camera Gain &   2.5 elec DN$^{-1}$ & Roll & 0.985$^\circ$ clockwise \\
Camera Noise*: & & Exposure Time & 2\,s\\
\hspace{0.2in}NE Quad & 4.0 DN & Full Data Set: &\\
\hspace{0.2in}NW Quad & 3.4 DN & \hspace{0.2in}No. of Images & 78\\
\hspace{0.2in}SE Quad  & 5.5 DN & \hspace{0.2in}Cadence & 4.4\,s\\
\hspace{0.2in}SW Quad  & 3.6 DN & Low Jitter Set: &\\
Plate Scale  & 0.129\arcsec\ pixel$^{-1}$ & \hspace{0.2in}No. of Images & 36 
\\
Resolution  & \textless\,0.47\arcsec (low jitter & \hspace{0.2in}Cadence & 4.4\,s \\ 
 &images only)  & & (periodic $\sim$20\,s gaps) \\ \hline
\end{tabular}
\label{tab:data_info}
\end{table}
\end{center}

\begin{figure}[H]
\begin{center}
\includegraphics[width=1.0\textwidth]{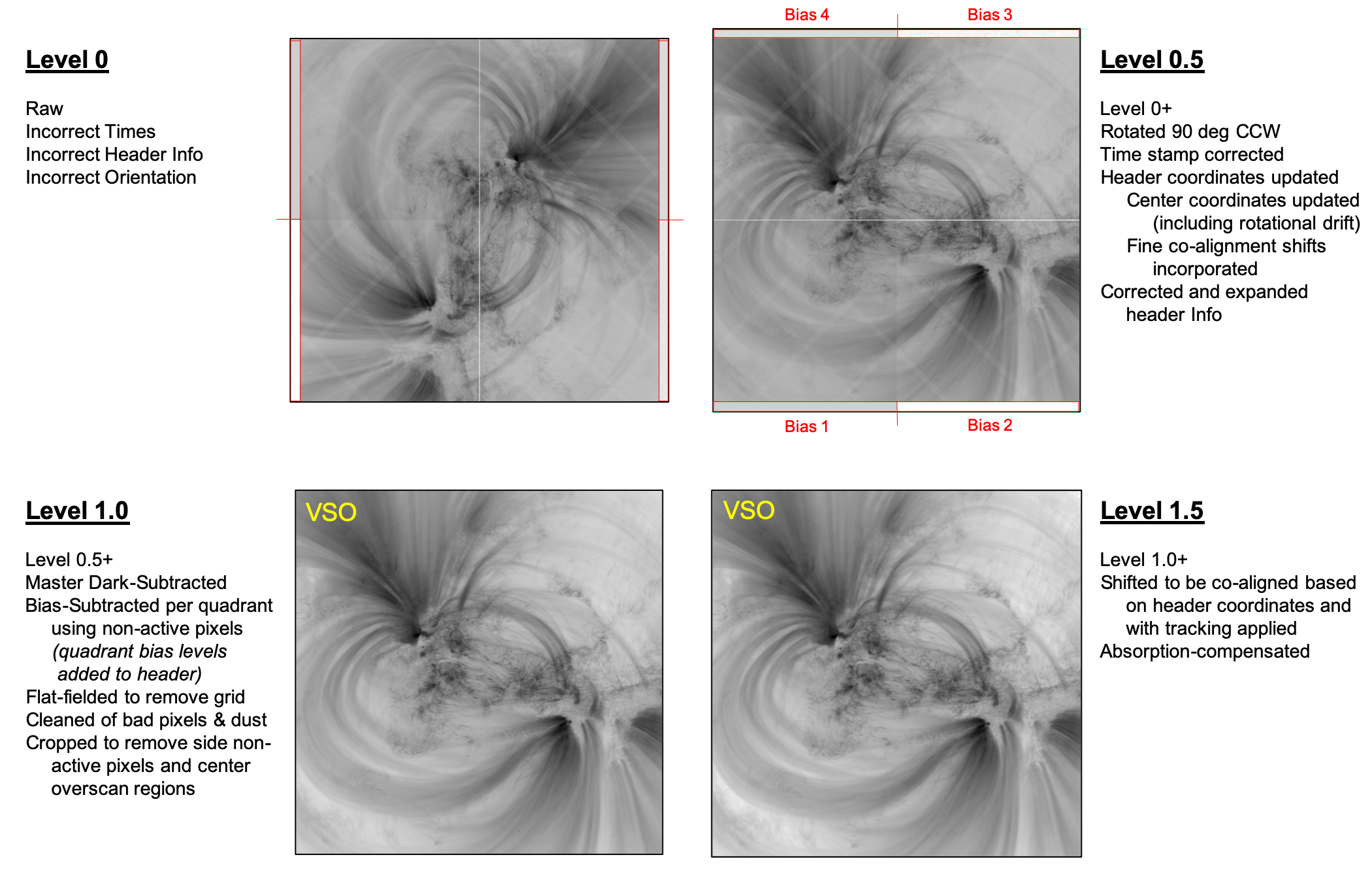}
\caption{Summary of the processed data levels and availability through the VSO.}
\label{fig:levels}
\end{center}
\end{figure}

Data sets have been generated at progressive levels of processing (see Figure~\ref{fig:levels}). Level~0.5 data was rotated to place solar north along the top of each frame, and the FITS file header information was updated (e.g., corrected time and pointing).  Level~1.0 data is bias- and dark-subtracted, flat-fielded, bad pixel-corrected, and cropped to remove non-active and overscan regions.  Level~1.5 data is shifted to be co-aligned (with target tracking applied), and the intensity levels are compensated for atmospheric absorption. 

Levels~1.0 and 1.5 have been distributed by the science team via the Virtual Solar Observatory (VSO). Additional information on Hi-C\,2.1 image processing is available in the User Guide distributed with the data. For these two levels, an accompanying low jitter set is provided which excludes the frames most affected by motion blur from the rocket. 

\subsection{Co-observations} \label{sec:coobs}

The Hi-C\,2.1 flight was coordinated with several other ground- and space-based solar observing observatories, including IRIS, three telescopes onboard the \textit{Hinode} satellite, SDO/AIA, the National Solar Observatory Interferometric BIdimensional Spectropolarimeter (NSO/IBIS), the Nuclear Spectroscopic Telescope Array (NuSTAR), the Big Bear Solar Observatory, the Owens Valley Radio Observatory, and the Swedish Solar Telescope.  Due to various circumstances, particularly regarding weather, the most successful coordinations resulted from IRIS, the \textit{Hinode} suite, AIA, IBIS, and NuSTAR.

\begin{figure}[H]
\begin{center}
\includegraphics[width=1.0\textwidth]{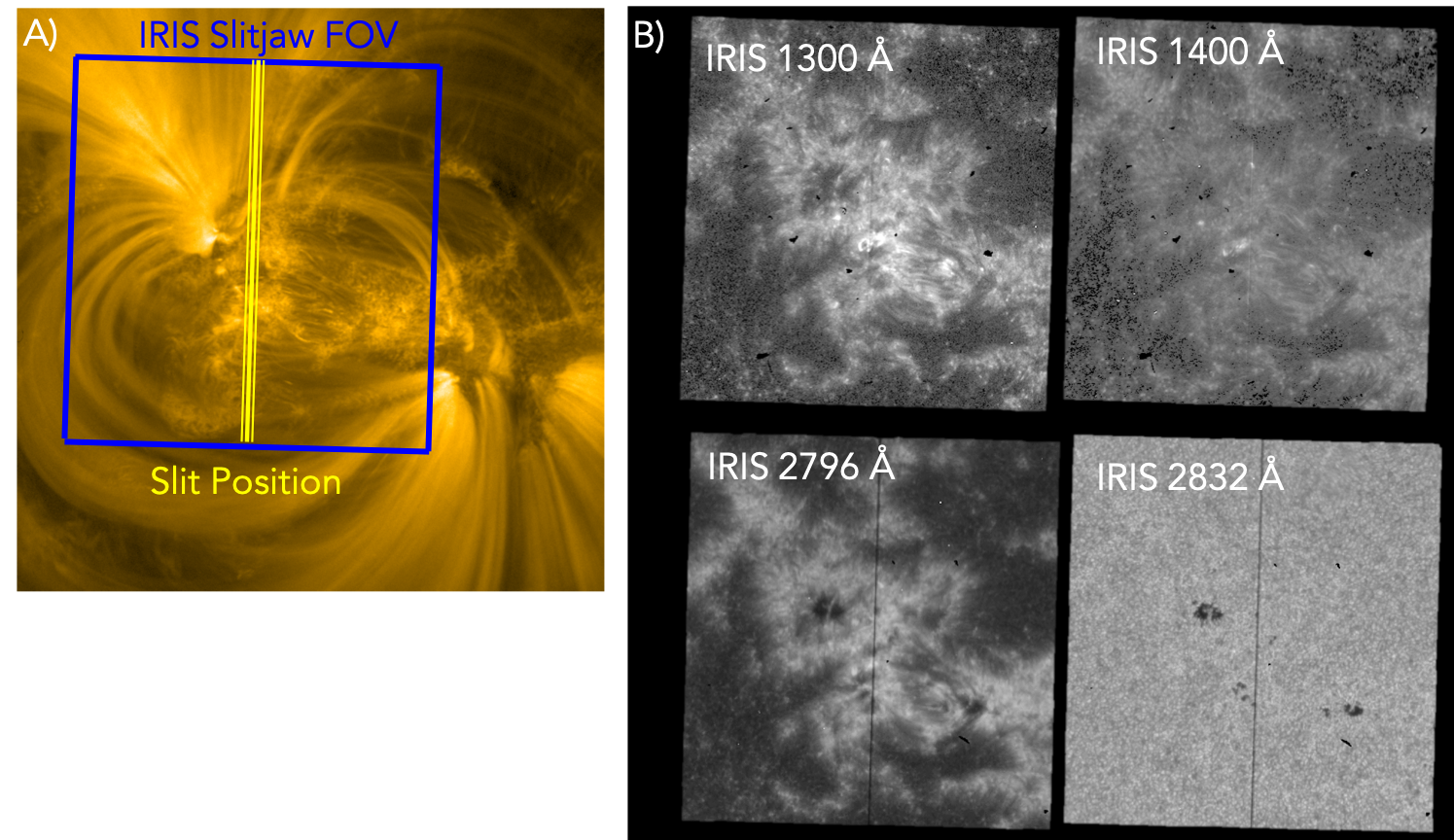}
\caption{Co-temporal IRIS observations. A) IRIS slitjaw FOV and slit position overlaid onto the Hi-C\,2.1 FOV. B) Example IRIS slitjaw images from the coordinated set.}
\label{fig:hic_iris}
\end{center}
\end{figure}

The primary science driver to connect chromospheric and coronal heating events necessitated key overlapping data from IRIS.  Fortunately, coordinations with IRIS were highly successful (IRIS OBSID 3600104031), providing high-resolution images and spectra of the AR core and loops within the Hi-C\,2.1 FOV (Figure~\ref{fig:hic_iris}). 

\begin{figure}[H]
\begin{center}
\includegraphics[width=1.0\textwidth]{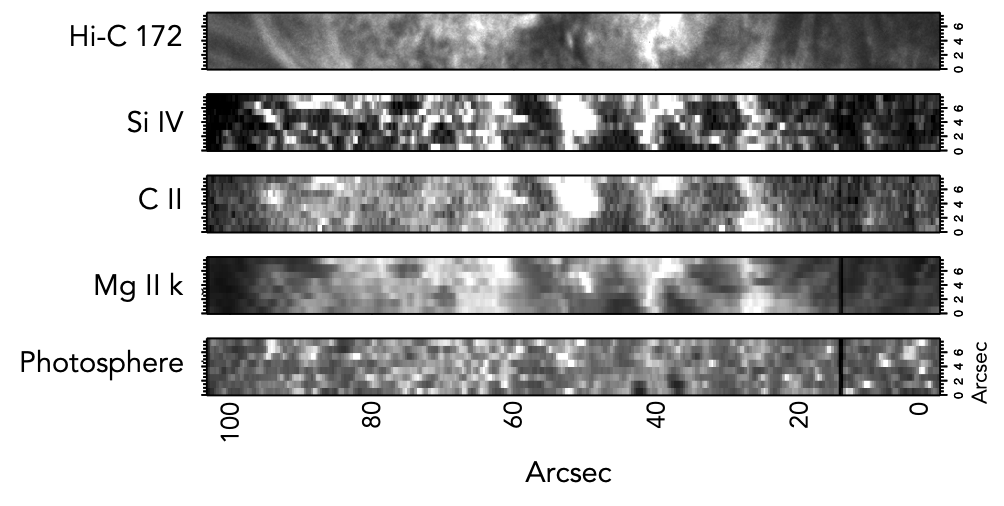}
\caption{Sparse IRIS rasters, summed spatially and spectrally by 2, of the photosphere, Mg\,II\,k core, C\,II, and Si\,IV lines along with the Hi-C\,2.1 intensity extracted to match the 8\,arcsec width of the IRIS raster.  Full spectral information of the IRIS line profiles is available at each spatial pixel.}
\label{fig:hic_iris_rasters}
\end{center}
\end{figure}

The IRIS spectroheliograms (Figure~\ref{fig:hic_iris_rasters}) were optimized to obtain fast rasters of a region of 8\arcsec\,x\,175\arcsec\ within 25\,s (using exposure times of 2\,s and sparse 1\arcsec\ steps).  To maximize faint transition region signals, the data was summed spatially and spectrally by 2, so that the resulting resolution is 0.66\arcsec\ along the slit (sampled at 0.33\arcsec) with 1\arcsec\ steps across the slit. At each slit location, spectral information is available for the C II 1335\AA, O I 1356\AA, Si IV 1394\AA, Si IV 1042\AA, and Mg II 2796\AA\,k and Mg II 2803\AA\,h lines. This unique combined dataset provides the first sub-arcsecond resolution dataset covering the full solar atmosphere from the photosphere, through the chromosphere and transition region into the corona.

\section{Conclusions} \label{sec:concl}
The third flight of the Hi-C sounding rocket, Hi-C\,2.1, has resulted in the highest resolution data of the cool corona and transition region ever taken at $172$\,\AA. The full set of calibrated data from the flight are now available on the VSO. Together with the coordinated observations, this data set provides a unique view of this important region, which will enhance our scientific knowledge of the coupling of mass and energy between the chromosphere and the corona. Despite the fact that the data only span $\sim 5$ minutes in time, the scientific advancements made by the first flight, and the data analysis already begun on this second set of data prove that the fine-scale structure of this region is key to our understanding of the solar atmosphere.

 \begin{acks}
We acknowledge the High-resolution Coronal Imager (Hi-C\,2.1) instrument team for making the second re-flight data available under NASA Heliophysics Technology and Instrument Development for Science (HTIDS) Low Cost Access to Space (LCAS) program (proposals  HTIDS14\textunderscore 2-0048 and HTIDS17\textunderscore 2-0033). MSFC/NASA led the mission with partners including the Smithsonian Astrophysical Observatory, the University of Central Lancashire, and Lockheed Martin Solar and Astrophysics Laboratory.  Hi-C\,2.1 was launched out of the White Sands Missile Range on 2018 May 29. The AIA data are courtesy of NASA/SDO and the AIA Science Team. S.K.T. gratefully acknowledges support by NASA contracts NNG09FA40C (IRIS), and NNM07AA01C (Hinode). The work of DHB was performed under contract to the Naval Research Laboratory and was funded by the NASA Hinode program.
 \end{acks}

%
%
\bibliographystyle{spr-mp-sola}
\bibliography{hic21_bib}  
%
%
%
%

\end{article} 
\end{document}